\documentclass[aps,prd,amsmath,amssymb,amsfonts,floats,floatfix,superscriptaddress,nofootinbib,notitlepage]{revtex4}
\usepackage{graphicx}
\usepackage{graphics}
\usepackage{color}
\usepackage{xspace}

\begin{document}

\newcommand{\fixme}[1]{\textcolor{red}{[#1]}}

\newcommand{\normnbd}{\cal{N}}
\newcommand{\nn}{\nonumber}
\newcommand{\tauep}{\tau_\epsilon}
\newcommand{\alep}{\alpha_\epsilon}
\newcommand{\ep}{{\epsilon}}
\newcommand{\co}{{\cal{O}}}
\newcommand{\cg}{{\cal{G}}}
\newcommand{\cf}{{\cal{F}}}
\newcommand{\al}{\alpha}
\newcommand{\mt}{\mathbb{M}_2}
\newcommand{\st}{{\mathbb{S}^2}}
\newcommand{\dt}{{\Delta t}}
\newcommand{\xp}{{x^\prime}}
\newcommand{\xpp}{{x^{\prime\prime}}}
\newcommand{\tpp}{{t^{\prime\prime}}}
\newcommand{\tp}{{t^\prime}}
\newcommand{\yp}{{y^\prime}}
\newcommand{\vp}{{v^\prime}}
\newcommand{\ypp}{{y^{\prime\prime}}}
\newcommand{\thp}{{\theta^\prime}}
\newcommand{\thpp}{{\theta^{\prime\prime}}}
\newcommand{\phip}{{\phi^\prime}}
\newcommand{\sigt}{\Sigma_t}
\newcommand{\sigtp}{{\Sigma_{t^\prime}}}
\newcommand{\sigi}{{\sigma_1}}
\newcommand{\sigii}{{\sigma_2}}
\newcommand{\be}{\begin{equation}}
\newcommand{\ee}{\end{equation}}
\newcommand{\beq}{\begin{eqnarray}}
\newcommand{\eeq}{\end{eqnarray}}
\newcommand{\bes}{\begin{eqnarray*}}
\newcommand{\ees}{\end{eqnarray*}}
\newcommand{\hpi}{\hat{\pi}}
\newcommand{\tiltau}{\tilde{\tau}}
\newcommand{\edoc}{\end{document}}
\newcommand{\re}{\mathbb{R}}
\newcommand{\etae}{\etaM_{\epsilon}}
\newcommand{\etaM}{\eta}
\newcommand{\dgN}{\Delta\gamma_N}
\newcommand{\sigNe}{\sigma_N^{\epsilon}}
\newcommand{\ind}{n}
\newcommand{\lh}{(\ell+1/2)}
\newcommand{\lsumo}{\sum_{\ell=0}^\infty}
\newcommand{\lsumi}{\sum_{\ell=1}^\infty}
\newcommand{\plg}{P_\ell(\cos\gamma)}
\newcommand{\plgt}{\tilde{P}_\ell(\cos\gamma)}
\newcommand{\plgti}{\tilde{P}_\ell^{(1)}(\cos\gamma)}
\newcommand{\plgtii}{\tilde{P}_\ell^{(2)}(\cos\gamma)}
\newcommand{\plgtj}{\tilde{P}_\ell^{(j)}(\cos\gamma)}
\newcommand{\tilt}{\tilde{T}}
\newcommand{\ca}{{\cal{A}}}
\newcommand{\cb}{{\cal{B}}}
\newcommand{\cd}{{\cal{D}}}
\newcommand{\cu}{{\cal{U}}}
\newcommand{\grsing}{G_R^{\textrm{sing}}}

\title{A Kirchhoff integral approach to the calculation of Green's functions beyond the normal neighbourhood.}
\author{Marc Casals}
\email{marc.casals@ucd.ie;mcasals@perimeterinstitute.ca}
\affiliation{Perimeter Institute for Theoretical Physics, Waterloo, Ontario, Canada N2L 2Y5}
\affiliation{Department of Physics,
University of Guelph,
Guelph, Ontario, Canada N1G 2W1}
\affiliation{School of Mathematical Sciences and Complex \& Adaptive Systems
Laboratory, University College Dublin, Belfield, Dublin 4, Ireland}

\author{Brien C. Nolan}
\email{brien.nolan@dcu.ie}
\affiliation{School of Mathematical Sciences, Dublin City
University, Glasnevin, Dublin 9, Ireland.}

\begin{abstract}
We propose a new method for investigating the global properties of the retarded Green's function $G_R(\xp,x)$ for fields propagating on an arbitrary
globally hyperbolic spacetime. Our method combines the Hadamard form for $G_R$ (this form is only valid within a normal neighbourhood of $x$)
together with Kirchhoff's integral representation for the field in order to calculate $G_R$ outside the maximal normal neighbourhood of $x$.
As an example, we apply this method  to the case of a  scalar field on a black hole toy-model spacetime, the Pleba{\'n}ski-Hacyan spacetime, $\mt\times\st$. The method allows us to determine in an exact manner that the singularity structure of the `direct' term in the Hadamard form for
$G_R(\xp,x)$ changes from a form $\delta(\sigma)$ to `$1/\pi\sigma$' after the null geodesic joining $x$ and $x'$ has crossed a caustic point, where $\sigma$ is the world function. Furthermore, there is a change of form from a $\theta(-\sigma)$ to a `$-\ln|\sigma|/\pi$' in the `tail' term, which has not been explicitly noted before in the literature. We complement the results from the Kirchhoff integral method with an analysis for large-$\ell$ of the Green function modes. This analysis allows us to determine the singularity structure after null geodesics have crossed an arbitrary number of caustics, although it raises a causality issue which the Kirchhoff integral method resolves. Because of the similarity in the caustic structure of the spacetimes, we expect our main results for wave propagation to also be valid on Schwarzschild spacetime.
\end{abstract}
\pacs{04.70.-s, 04.62.+v} \maketitle



\section{Introduction}

Understanding the propagation of waves on a curved spacetime is of interest on its own right, and also has many important applications in various
physical settings.
Consider, for example, the inspiral of a black hole binary in the extreme mass ratio limit. This system
can be modelled by the motion of the smaller black hole in the gravitational field of the larger black hole (typically Schwarzschild or Kerr black hole)
under the action of a self-force~\cite{Poisson:2011nh} induced by its own gravitational field. At the heart of the analysis of such systems are the gravitational wave perturbations generated by the small black hole in the curved spacetime of the large black hole. In particular, the self-force is governed by an equation which involves the integral of the gradient of the retarded Green's function
of the corresponding wave equation.
The integral is taken over the entire past of the world line
of the particle.
While the local structure of the Green's function is well known in terms of the so-called Hadamard form~\cite{DeWitt:1960},
a lot less is known about the {\it global} structure of the Green's function.
Clearly, however, the structure - in particular, the singularity structure - of the Green's function for spacetime points arbitrarily separated is crucial for the understanding of wave propagation and it can be of practical use for the calculation of the self-force. This is the basic aim of the present paper: to introduce a method that allows the calculation of the retarded Green's function at arbitrarily separated spacetime points. We apply this method to determine new results regarding the singularity structure of the retarded Green's function in a black hole toy-model spacetime.

It is known that the singularities of the Green's function occur whenever the two spacetime
points are connected by a null geodesic (see e.g.\ \cite{Garabedian, Ikawa}).
Within a normal neighbourhood of the base point $x$ (i.e. a region $\normnbd$$(x)$ containing $x$ with the property that every $\xp\in\normnbd$$(x)$ is connected to $x$ by a unique geodesic which lies in $\normnbd$$(x)$),
the Hadamard form dictates that the singularity of the retarded Green function is of the type $\delta(\sigma)$, where $\sigma=\sigma(x,x')$ is Synge's world-function,
i.e., one half of the geodesic distance along the unique geodesic connecting the points $x$ and $\xp$.
In principle, outside a normal neighbourhood the biscalar $\sigma(x,x')$ is not well-defined and the Hamadard form is not valid; therefore,
one must take an alternative approach to the calculation of the Green function.
In~\cite{CDOWa} it was shown by using a quasinormal mode series
that the scalar Green's function $G_R(x,x')$ presents a four-fold singularity structure:
$\delta(\sigma), \text{P.V.}\left(1/\pi\sigma\right), -\delta(\sigma), - \text{P.V.}\left(1/\pi\sigma\right), \delta(\sigma), \dots$, where the change of character in the singularity occurs every time
that the null geodesic connecting the two points passes through a caustic (i.e., a spacetime point where neighbouring null geodesics are focused).
This was shown specifically for the case of a static region of Nariai spacetime ($dS_2\times\st$) but it was anticipated to be also valid in most situations in spherically symmetric spacetimes. (We anticipate exceptional points in every spherically symmetric spacetime - see Eq.(\ref{GRpires}) - and indeed there are exceptional spacetimes, for example Bertotti-Robinson spacetime.)
Such a four-fold structure in General Relativity had been previously noted by Ori~\cite{Ori1short}.
It was subsequently shown in~\cite{Dolan:2011fh}, again by using a quasinormal mode series, that this four-fold structure is also present in Schwarzschild spacetime --
although, in this case, the singularity times given by a (truncated) quasinormal mode series only approximate
the singularity times given by a null geodesic connecting the two spacetime
points.

Furthermore, very recently it has been shown in~\cite{Harte:2012uw} that it is possible to probe the global singularity structure of the Green's function on a generic spacetime by exploiting the Penrose limit \cite{Penrose} and results relating singularity structures and caustics in plane wave spacetimes. In particular, the authors find that the four-fold singularity structure applies in a wide class of spacetimes including Schwarzschild (and most likely, Kerr).


A causality issue is raised by the presence of the singular term $\text{P.V.}\left(1/\pi\sigma\right)$ in its naive form indicated above:
it raises a  question regarding the causal character of the Green's function, since $\text{P.V.}\left(1/\pi\sigma\right)$ is nonzero even when the spacetime points are spacelike separated. While the quasinormal mode method is of practical application (albeit to special choices of the mass and, in non-Ricci-flat spacetimes, of the coupling constant), it does not answer this  question and it does not generally offer exact control of the terms neglected. The same holds for the Penrose limit approach of \cite{Harte:2012uw}. This issue is resolved in our approach, which also has the advantage of deriving a corresponding four-fold structure for the so-called tail term of the retarded Green's function. We note in particular that this term contributes to the singular part of the Green's function after the formation of (an odd number of) caustics.

The normal neighbourhood, and hence also the Hadamard form for $G_R$,
generally breaks down at caustic points of the background spacetime.
In a spherically symmetric spacetime,
the spray of null geodesics emitted from an event $x$
can include an $\st$-envelope that reconverges at a point $\xp$ -- a caustic -- whose angular separation from $x$ is $\gamma=\pi$.
It seems to be precisely the focusing of this $\st$-envelope of null geodesics that yields the four-fold singularity structure in $G_R(x,x')$ for points with $\gamma \neq 0,\pi$. (For example, in the case of the Einstein Static Universe, $\mathbb{R}\times \mathbb{S}^3$, the singularity structure is two-fold, and of course the singularity structure is uniform in flat spacetime.)
This is the case in Schwarzschild spacetime, but it is also the case in a spacetime $M=M_2\times \st$,
the direct product of a 2-dimensional spacetime $M_2$ of constant curvature
and the unit 2-sphere $\st$ (both equipped with their standard metrics).
Such spacetimes have found application in different contexts in general relativity (see section 7.2 of~\cite{Griffiths&Podolsky}).

In this paper we develop a new method for the formal construction of the retarded Green's function that applies beyond the normal neighbourhood. Consider a field $\phi$ propagating on a globally hyperbolic spacetime, whose evolution is governed by a homogeneous wave equation.
Kirchhoff's formula~\cite{Poisson:2011nh} gives a representation of $\phi$ at a certain point $x$ as a convolution integral of $G_R$, the retarded Green's function of the wave equation, and its derivative, with initial data for $\phi$ on
a spacelike hypersurface in the past of $x$. In the Kirchhoff integral, $G_R$ acts as a propagator for the initial data for $\phi$. By noting that $G_R(x,\xp)$ satisfies the {\it homogeneous} wave equation if $\xp$ lies strictly to the future
of $x$, we can apply Kirchhoff's formula to the retarded Green's function itself instead of the field.
In our method, which we refer to as the Kirchhoff integral method, we apply Kirchhoff's formula with $G_R$ acting both as propagator and data. We choose time separations so that the Hamadard form is valid for both versions of $G_R$ appearing in the Kirchhoff (propagator and data): this puts an upper bound on the time step we can take. However, through repeated application of Kirchhoff's formula in this manner we can, in theory, calculate $G_R$ for points {\it arbitrarily} separated with the sole knowledge of the Hadamard form for $G_R$.

We use this method to study the singularity structure of the Green's function as the propagating field passes through caustic points of the background
spacetime.
We note that in~\cite{DeWitt:1964de} a method was employed which has certain similarities with
our Kirchhoff integral method, although they are fundamentally different methods. The Kirchhoff integral method also essentially underpins the standard derivation of the reciprocity relationship $G_R(x,\xp)=G_A(\xp,x)$ between the retarded and advanced Green's functions~\cite{M&F,Poisson:2011nh}, and to determine properties of the Detweiler-Whiting singular Green's function: see Section 14.5 of \cite{Poisson:2011nh}.
Our method is in principle valid for any globally-hyperbolic spacetime.
However, in practise one must have a knowledge of the Hadamard form for $G_R$.

In this paper, we apply the Kirchhoff integral method to the case of a scalar field propagating in a black hole toy-model spacetime
$M=\mt\times \st$, where
$\mt$ is 2-dimensional Minkowski spacetime.
Following~\cite{Griffiths&Podolsky}, we will refer to this spacetime $M$ as Pleba\'nski-Hacyan (PH) spacetime.
We investigate properties of wave propagation on PH in the case of a scalar field created by a scalar charge. However, many of the properties of wave propagation are independent of the spin of the propagating field and we believe that the approach presented here would also be valid in the higher-spin cases.

Given that the calculations are rather involved, we present here the main result that we obtain when applying the Kirchhoff integral method to PH.
For field points $\xp$ in a normal neighbourhood of the base point $x$, the \textit{Hadamard form} for the retarded Green's function applies. This is given
by~\cite{DeWitt:1960,Friedlander,Poisson:2011nh}
\be G_R(x,\xp)= [U(x,\xp)\delta(\sigma(x,\xp))+V(x,\xp)H(-\sigma(x,\xp))]H_+(x,\xp),\label{grhad}\ee
where $\delta,H$ are the usual Dirac delta and Heaviside distributions,
\[ H_+(x,\xp)=\left\{\begin{array}{ll} 1 & \hbox{if $x$ lies to the future of $\xp$};\\ 0 & \hbox{otherwise},\end{array}\right. \]
and $U, V$ are biscalars determined as follows. It can be shown that
$U(x,\xp)=\Delta^{1/2}(x,\xp)$,
where $\Delta(x,\xp)$ is the van Vleck-Morette determinant (which again, is defined only when $\xp$ lies in a normal neighbourhood of $x$).
The bitensor $V(x,\xp)$ can be written as an asymptotic series, $V(x,x')=\sum_{n=0}^\infty \nu_n(x,x')\sigma^n$, where $\nu_n(x,x')$ are coefficients satisfying
certain recurrence relations~\cite{DeWitt:1960,Decanini:Folacci}.
We refer to the terms $U\delta(\sigma)$ and $VH(-\sigma)$ in the Hadamard form (\ref{grhad}) as, respectively, the
`direct' and the `tail' terms.

The world function of PH spacetime can be written as $\sigma = -\frac12\eta^2 + \frac12\gamma^2$ where $\eta$ is the geodesic distance on $\mt$ and $\gamma$ is the geodesic distance on $\st$. Then for points $\xp$ in a normal neighbourhood of $x$, the Hadamard form applies with
\[ U(x,\xp) = U(\gamma) = \left|\frac{\gamma}{\sin\gamma}\right|^{1/2},\qquad V(x,\xp)=\sum_{k=0}^\infty \nu_k(\gamma)\sigma^k.\]
where $\nu_k$ are smooth functions of $\gamma$ satisfying certain recurrence relations.
Notice then that we can write
$V(x,\xp)=\Upsilon(\eta,\gamma)$.
The normal neighbourhood condition breaks down at $\eta=\pi$, corresponding to the formation of the first caustic.
The second caustic forms at $\eta=2\pi$. Our main result from the application of the Kirchhoff integral method to PH is that, for $\pi<\eta<2\pi$,
\begin{eqnarray} G_R(x,\xp)= U(2\pi-\gamma)PV\left(\frac{1}{\pi\sigma_1(x,\xp)}\right)-\nu_0(2\pi-\gamma)\frac{\ln|\sigma_1(x,\xp)|}{\pi}+\co(1),
\label{GRresult}
\end{eqnarray}
where
\be \sigma_1 = -\frac12\eta^2+\frac12(2\pi-\gamma)^2,\label{sigma1forGR}\ee
and the $\co(1)$ term corresponds to a function that remains finite throughout the region $\pi<\eta<2\pi$. We note that
\[ \nu_0(2\pi-\gamma)= \Upsilon(\eta,2\pi-\gamma)|_{\sigma_1=0}.\]
That is, we show that, after the wavefront has crossed one caustic,
the `direct' term in the Green's function changes its singularity character from $\delta(\sigma)$ to
$\text{P.V.}\left(1/\pi\sigma\right)$, as expected, and that the `tail' term gives rise to a singular term $\ln|\sigma|$ in the same region.
This provides the complete description of the singular part of the retarded Green's function after the wavefront has crossed one caustic point.
To the best of our knowledge, the diverging term $\ln|\sigma_1|$ has not been noted before explicitly in any spacetime.
We expect a similar singularity to occur in Schwarzschild spacetime.

We note that the four-fold singularity structure degenerates to a two-fold singularity structure at the caustic point itself, which entails spacetime points with angular separation $\gamma=\pi$. This is demonstrated explicitly below: see Section IV-F and especially Eq.(\ref{GRpires}). The dominant part of the singularity displays the form $\delta(\sigma)\to-\delta(\sigma)$ at $\gamma=\pi$.

Our method answers cleanly the two questions raised above, the one about an extension of the world function $\sigma$ outside the normal
neighbourhood (here given by $\sigma_1$ in (\ref{sigma1forGR}))
and the one about `acausality' due to $\text{P.V.}\left(1/\pi\sigma\right)$ (the result (\ref{GRresult}) here
is only valid for $\pi<\eta<2\pi$),
by treating {\it directly} the spacetime in question and obtaining {\it exact} results.
Furthermore, we clearly identify the coefficients of the $PV$ and $\ln$ singularities, suggestively expressed in terms of the bitensors $U$ and $\nu_0$ defined in a normal neighbourhood of $x$.

We complement the new Kirchhoff integral method with a large-$\ell$ asymptotic analysis of the Green's function modes
in PH spacetime, where $\ell$ is the multipole number. This analysis is very similar in nature to the quasinormal mode method used in~\cite{CDOWa,Dolan:2011fh} (although here it is not necessary to perform a Fourier transform and so the modes correspond to a multipole decomposition only).
As such, and unlike the Kirchhoff integral method, the large-$\ell$ asymptotic analysis
uncovers the singularity structure for an {\it arbitrary} number of caustic crossings, but it does not resolve the causality issue regarding
the term $\text{P.V.}\left(1/\pi\sigma\right)$ mentioned above.
The large-$\ell$ analysis that we carry out here  yields not only the four-fold structure for the `direct' term ($\pm\delta(\sigma)\to \mp\text{P.V.}\left(1/\pi\sigma\right)$) but also that for the `tail' term ($\pm\theta(-\sigma) \to\mp \ln\left|\sigma\right|/\pi$), in both cases for any number of caustic crossings.
Fig.\ref{fig:Kirchhofff unwrapped cylinder} shows the four-fold structure as obtained from an $\ell$-mode series.
A shortcoming of the large-$\ell$ asymptotic analysis is that we must restrict to a particular value of $m^2+\xi R$, where $m$ is the mass, $\xi$ is the coupling term and $R$ is the Ricci scalar.

The layout of the paper is as follows. 
In Sec.\ref{sec:Kirchhoff}, we describe the Kirchhoff integral method as can be applied to a general, globally-hyperbolic spacetime.
In Sec.\ref{sec:M1xM2} we consider briefly the world function and the Hadamard form for $G_R$ in spacetimes of the type $M_1\times M_2$ and, in particular, in PH spacetime.
In Sec.\ref{sec:Kirchhoff in PH} we apply the Kirchhoff integral method specifically to the PH spacetime and provide the lengthy and detailed calculations leading to the result (\ref{GRresult}). This long section is almost entirely technical in nature, and stands independent of the remainder of the paper except insofar as it provides the proof of our main result. The large-$\ell$ calculation is given in Sec.\ref{sec:Large l} and we conclude with some final remarks in Sec.\ref{sec:conclusions}.



\section{Kirchhoff Method for the Retarded Green's Function} \label{sec:Kirchhoff}

We consider the wave equation for a scalar field $\Psi(x)$,
\be (\Box-m^2-\xi R)\Psi(x)=0,\label{wave}\ee
where $\Box$ is the d'Alembertian operator, $m$ is the scalar mass, $R$ is the Ricci scalar and $\xi$ is a coupling constant.
The retarded Green's function for this equation satisfies
\be (\Box-m^2-\xi R)G_R(x,\xp)=-4\pi\delta_4(x,\xp),\label{weGret}\ee along with
the boundary conditions $G_{ret}(x'',x)=0$ if $x''\notin J^+(x)$. The quantity $\delta_4$ is the 4-dimensional Dirac distribution.
A useful property of $G_R$ is that it allows us to explicitly write down the value of a field $\Psi$ at a point $x$ to the future of
an initial data hypersurface $\Sigma^\prime$~\cite{Poisson:2011nh,DeWitt:1960}:
\be \Psi(x)=-\frac{1}{4\pi}\int_{\Sigma^\prime}(G_R(x,\xp)\nabla^{\alpha^\prime}\Psi(\xp)-\Psi(\xp)\nabla^{\alpha^\prime}G_R(x,\xp))d\Sigma^\prime_{\alpha^\prime},\label{kirchhoff}\ee
where $d\Sigma^\prime_{\alpha^\prime}=-n_{\alpha^\prime}dV^\prime$ is the surface element on $\Sigma^\prime$, $n_{\alpha^\prime}$ the future directed normal one-form and $dV^\prime$ is the invariant volume element on $\Sigma^\prime$. Thus the \textit{Kirchhoff integral} (\ref{kirchhoff}) involves the convolution of $G_R$ and its derivative with the initial data for $\Psi$ on $\Sigma^\prime$.




Consider now a globally hyperbolic spacetime $(M,g_{\alpha\beta})$ with a global time function $T$. Then the hypersurfaces
\[ \sigt=\{x\in M:T(x)=t\}\] are Cauchy surfaces for the spacetime.
At this point, we implement the following notational change. We take $\xpp$ to be the base point for our calculations, and consider points on hypersurfaces $\sigtp$ and $\Sigma_t$ that lie to the future of the hypersurface $\Sigma_\tpp$. So with
\[ T(x)=t>T(\xp)=\tp>T(\xpp)=\tpp,\] we see that the retarded Green's function $G_R(x,\xpp)$ satisfies the \textit{homogeneous} wave equation in the $x$ variables, and so we can apply the Kirchhoff integral (\ref{kirchhoff}) to write down
\be G_R(x,\xpp)=\frac{1}{4\pi}\int_{\sigtp}(G_R(x,\xp)\partial_{\tp}G_R(\xp,\xpp)-\partial_{\tp}G_R(x,\xp)G_R(\xp,\xpp))dV^\prime.\label{kirch}\ee
In this integral, $G_R(x,\xp)$ acts as the propagator for the field $G_R(\xp,\xpp)$, carrying it from $\sigtp$ to $\Sigma_t$. We note that (\ref{kirch}) is well-defined as a distribution, being the convolution of distributions with compact support: see sections 2.5 and 2.9 of~\cite{Friedlander}.

The formula (\ref{kirch}) is of particular interest when the time intervals $\tp-\tpp$ and $t-\tp$  are such that both $G_R(\xp,\xpp)$ and $G_R(x,\xp)$ may be written in Hadamard form, but the field point $x$ \textit{does not} lie in a normal neighbourhood of the base point $\xpp$. Hence (\ref{kirch}) (formally) provides a closed form for the retarded Green's function beyond the normal neighbourhood. By applying (\ref{kirch}) iteratively in this way, we have a mechanism for calculating $G_R$ globally.



\section{Product spacetimes $M_1\times M_2$} \label{sec:M1xM2}


\subsection{General case}
Let us consider a product manifold $M=M_1\times M_2$, i.e.,
$ds^2=ds_1^2+ds_2^2$ where $ds_1^2=g_{AB}dx^Adx^B$ and $ds_2^2=g_{ab}dx^adx^b$.
Greek letters $\alpha, \beta$ run over all the coordinates in $M$, capital Latin letters $A,B$ run over the coordinates
in $M_1$ and small Latin letters $a,b$ over the coordinates in $M_2$.
Within this subsection only, subindices $1$ and $2$ will indicate quantities corresponding to the manifolds $M_1$ and $M_2$, respectively.
Then the defining property $\nabla_{\alpha}\sigma\nabla^{\alpha}\sigma=2\sigma$
implies that
\begin{eqnarray}
&\sigma=\sigma_1+\sigma_2,
\\
&\nabla_{A}\sigma_1\nabla^{A}\sigma_1=2\sigma_1,
\nonumber
\\
&\nabla_{a}\sigma_2\nabla^{a}\sigma_2=2\sigma_2
\nonumber
\end{eqnarray}
We now assume that $M_1$ is a 2-dimensional spacetime, $M_2$ is a 2-dimensional Riemannian manifold and so $M$ is a
4-dimensional spacetime.
Defining $\lambda $ and $\mu$ via $\sigma_1=-\lambda ^2/2$ and $\sigma_2= \mu ^2/2$ we then have
that
\begin{eqnarray}
&
\nabla_{A} \lambda\nabla^{A} \lambda =-1,\quad \Box_1\sigma_1=1-\lambda\Box_1\lambda
\\&
\nabla_{a} \mu\nabla^{a} \mu =+1,\quad \Box_2\sigma_2=1+\mu\Box_2\mu
\nonumber
\end{eqnarray}
These immediately imply that
\begin{eqnarray}
&
\nabla^{\alpha}\sigma\nabla_{\alpha}F(\lambda,\mu)=\lambda\dfrac{\partial F}{\partial\lambda}+ \mu\dfrac{\partial F}{\partial\mu}
\\&
\Box F(\lambda,\mu)=-\dfrac{\partial^2F}{\partial\lambda^2}+\dfrac{\partial F}{\partial\lambda}\Box_1\lambda
+\dfrac{\partial^2F}{\partial \mu ^2}+\dfrac{\partial F}{\partial \mu}\Box_2 \mu
\nonumber
\end{eqnarray}
for a sufficiently smooth function $F$ of $\lambda$ and $\mu$.
From the equation satisfied by the van Vleck determinant $\Delta(x,x')$, $\nabla_{\alpha}\left(\Delta\sigma^{\alpha}\right)=4\Delta$, we find
\begin{eqnarray}
&\Delta= \Delta_1\cdot \Delta_2, \label{vvmproduct}
\\&
\sigma_1^A\nabla_A\Delta_1+\Delta_1\Box_1\sigma_1=2\Delta_1
\nonumber\\&
\sigma_2^a\nabla_a\Delta_2+\Delta_2\Box_2\sigma_2=2\Delta_2
\nonumber
\end{eqnarray}

As shown in~\cite{Poisson:2011nh}, the restriction of the Hadamard biscalar $V(x,x')$ to the null cone $\hat{V}:= V|_{\sigma=0}=\nu_0|_{\sigma=0}$ satisfies a certain transport equation along the null geodesics ruling the null cone.
The bitensor $\nu_0(x,x')$ is defined by $V=\sum_{n=0}^\infty \nu_n\sigma^n$, as given in the introduction.
We can take $\Delta^{-1/2} \nu_0=x_0(\lambda)+y_0(\mu)$ to obtain
\begin{eqnarray}\label{eq:eqs x0,y0}
&
\dfrac{d(\lambda x_0)}{d\lambda}-f_1(\lambda)+\frac14(m^2+\xi R)=-\dfrac{d(\mu y_0)}{d\mu}-f_2(\mu)-\frac14(m^2+\xi R)=const.
\\ &
2f_1(\lambda)=\Delta_1^{-1/2}\left[-\dfrac{d^2\Delta_1^{1/2}}{d\lambda^2}+\Box_1\lambda\dfrac{d\Delta_1^{1/2}}{d\lambda}\right],
\quad
2f_2(\mu)=\Delta_2^{-1/2}\left[\dfrac{d^2\Delta_2^{1/2}}{d \mu ^2}+\Box_2 \mu\dfrac{d\Delta_2^{1/2}}{d \mu}\right],
\nonumber
\end{eqnarray}
where $f_1$ and $f_2$ are some functions of $\lambda$ and $\mu$, respectively.
It then follows that
\begin{equation}\label{eq hat V0}
\Delta^{-1/2} \nu_0=-\frac{\xi R}{2}+\int d\lambda\ f_1(\lambda)+\int d \mu\ f_2(\mu)+\frac{c_{x0}}{\lambda}+\frac{c_{y0}}{\mu}
\end{equation}
where $c_{x0}$ and $c_{y0}$ are constants of integration to be determined by the condition that $[\nu_0]=\left[V\right]=(1-3(m^2+\xi R(x))/6$,
where the brackets $[\ ]$ around a certain bitensor $A(x,x')$ indicate the coincidence limit of that bitensor (i.e., $\lim_{x\to x'}A(x,x')$).
This condition follows
from the transport equation that $\hat{V}$ satisfies.



\subsection{PH: $\mt\times \st$}
The line element of PH spacetime can be written as
\be ds^2=-dT^2+dY^2+d\Omega^2,\label{PHlel}\ee
where $(T,Y)$ are global inertial coordinates on $\mt$ and $d\Omega^2$ is the standard line element on $\st$:
\[d\Omega^2=d\Theta^2+\sin^2\Theta d\Phi^2\]
in standard polar and azimuthal coordinates $(\Theta,\Phi)$. The Ricci tensor $R_{\al\beta}$ and Ricci scalar satisfy
\[ R_{\al\beta}dx^\al dx^\beta = d\Omega^2,\qquad R=2\]
and the only non-vanishing Weyl curvature component can be written in Newman-Penrose form as
\[ \psi_2 = -\frac16.\]
The d'Alembertian is given by
\begin{equation}
\Box=\Box_{\mathbb{M}_2}+\Box_{\mathbb{S}^2}, \quad
\Box_{\mathbb{M}_2}=-\frac{\partial^2}{\partial t^2}+\frac{\partial^2}{\partial y^2},\quad
\Box_{\mathbb{S}^2}=\frac{\partial^2}{\partial \gamma^2}+\cot\gamma \frac{\partial}{\partial \gamma}=
\frac{1}{\sin\gamma}\frac{\partial}{\partial \gamma}\left(\sin\gamma\frac{\partial}{\partial \gamma}\right).
\end{equation}
We find that
\be \sigma(x,\xp)=-\frac12(t-\tp)^2+\frac12(y-\yp)^2+\frac12\gamma_1^2 \label{sigdef}\ee
where $t^\prime=T(\xp), \theta=\Theta(x)$ etc and
\be \cos\gamma_1=\cos\theta\cos\thp+\sin\theta\sin\thp\cos(\phi-\phip).\label{gamdef}\ee We write the world function on $\mt$ as
\be \bar{\sigma}(x,\xp)=-\frac12\eta^2=-\frac12(t-\tp)^2+\frac12(y-\yp)^2. \label{sigbardef}\ee
We note that when $\xp$ lies in a normal neighbourhood of $x$, $\sigma$ as given here is the world function of PH spacetime, but we will also understand (\ref{sigdef}) to serve as the global definition of a two point function on $\mt\times \st$.

In this spacetime, it is straightforward to show that the $\st$-envelope of null geodesics from any point $\xpp$ with $T(\xpp)=\tpp$ reconverges at a point $\xp$ with $T(\xp)=\tpp+\pi$. If $T(\xp)<T(\xpp)+\pi$, then $\xpp$ and $\xp$ are connected by at most one causal geodesic. Hence, in line with our comment at the end of Sec.\ref{sec:Kirchhoff}, the Kirchhoff formula (\ref{kirch}) is of most interest when
\be 0<\tp-\tpp<\pi \hbox{ and } 0<t-\tp<\pi\label{tsep1}\ee
and when
\be \pi<t-\tpp<2\pi. \label{tsep2}\ee

\begin{figure}[h!]
\begin{center}
\includegraphics[width=12cm]
{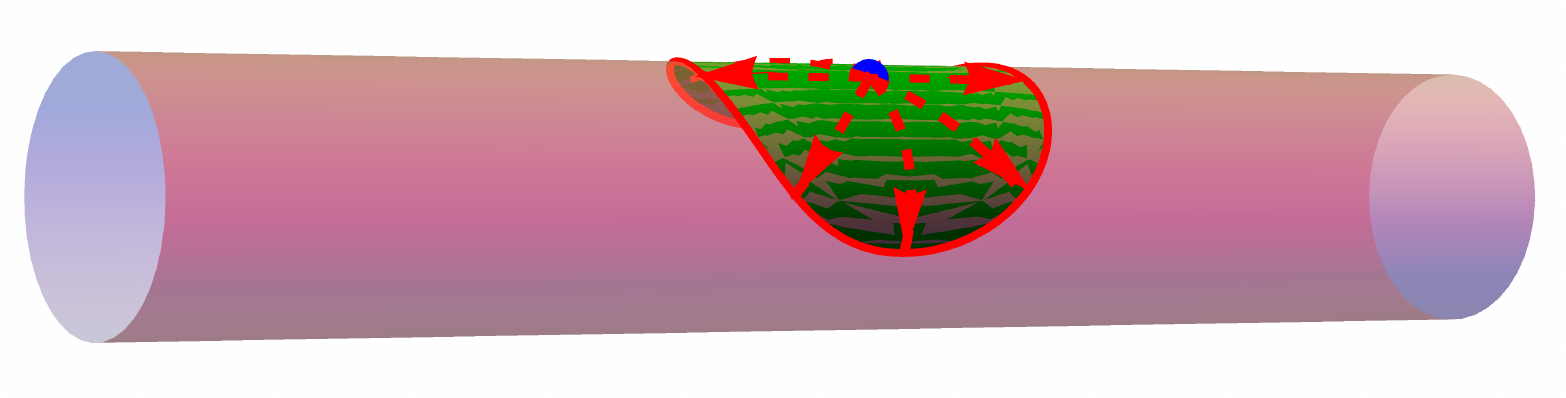}
\includegraphics[width=12cm]
{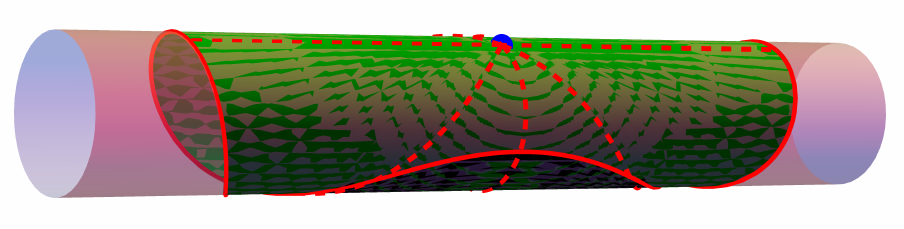}
\end{center}
\caption{
(Colour online.) Visualization of PH spacetime $\mathbb{M}_2\times \mathbb{S}^2$ at given spacelike slices $T=T_0=$ constant:  $T_0\in (0,\pi)$ in the top figure (referred to as (a) below)
and  $T_0\in (\pi,2\pi)$ in the bottom figure (b).
The coordinate along the axis of the cylinder corresponds to $Y$ and the angle around the cylinder corresponds to the azimuthal
angle $\Phi$;
the polar angular coordinate
$\Theta$ is suppressed. The blue dot is the base point $x$.
The continuous red line is the null wavefront, emitted at time $T=0$, at the time $T=T_0$.
The dashed red line corresponds to the trajectories up to time $T=T_0$ of representative null geodesics.
The region shaded in green corresponds to the intersection of the normal neighbourhood of $x$ with the hypersurface $\{T=T_0\}$.
In (b), the region shaded in black at the bottom of the figure corresponds to spatial points which, at the time $T=T_0$, have been reached by the wavefront
after crossing one caustic point. This region lies outside the maximal normal neighbourhood of the base point $x$.
}
\label{fig:M2xS2}
\end{figure}

Fig.\ref{fig:M2xS2} offers a visualization of null geodesics propagating on $\mathbb{M}_2\times \mathbb{S}^2$ on two different $T=const.$ slices.

We consider now the Hadamard form on PH spacetime. We can apply (\ref{vvmproduct})
to obtain
\begin{equation} \label{udef}
U(x,x')= U(\gamma)=
\left|\frac{\gamma}{\sin\gamma}\right|^{1/2}
\end{equation}

%
%

In a general spacetime, the coefficient $V(x,\xp)$ of the tail term in (\ref{grhad}) can be written in the form $V=\sum_{n=0}^\infty \nu_n\sigma^n$.
This is an asymptotic series for $V$ which converges uniformly in a normal neighbourhood of $x$ in a general class of spacetimes~\cite{DeWitt:1960,Friedlander}.
The coefficients are obtained by solving certain recurrence relations, which correspond to
a sequence of transport equations~\cite{DeWitt:1960,Decanini:Folacci}. In PH spacetime, we find that $\nu_k=\nu_k(\gamma)$ so that
 \be V=\sum_{k=0}^\infty \nu_k(\gamma)\sigma^k,\label{vdef}\ee
 and the transport equations become ordinary differential equations. Applying Eq.(\ref{eq hat V0}), we obtain
\be \hat{V} = V|_{\sigma=0}=\nu_0(\gamma)=\frac18U(\gamma)\left(1-4m^2-4\xi R+\frac{1}{\gamma^2}-\frac{\cot\gamma}{\gamma}\right).\label{v0def}\ee


%
%
%
%

Writing $\nu_k=U(\gamma)\tilde{\nu}_k(\gamma)$, the recurrence relations become
\begin{equation} \label{eq:rec rln Vk M2xS2}
\frac{d}{d\gamma}\left(\gamma^{k+1}\tilde{\nu}_k\right)=
-\frac{1}{2k}\frac{\gamma^{k-1/2}}{\sqrt{\sin\gamma}}\left\{
\frac{d}{d\gamma}\left[\sin\gamma\frac{d}{d\gamma}\left(\left(\frac{\gamma}{\sin\gamma}\right)^{1/2}\tilde{\nu}_{k-1}\right)\right]
-(m^2+\xi R)\sin\gamma\left(\frac{\gamma}{\sin\gamma}\right)^{1/2}\tilde{\nu}_{k-1}\right\}.
\end{equation}

From Eqs.(\ref{v0def}) and (\ref{eq:rec rln Vk M2xS2}) we find
\begin{equation} \label{eq:hat V1 M2xS2}
\tilde{\nu}_1=\frac{2\gamma^2-3 \csc ^2(\gamma ) \left[6 \gamma ^2+2 \gamma \sin (2 \gamma )+5 \cos (2 \gamma )-5\right]}{256 \gamma ^4}.
\end{equation}
for the specific case $m^2+\xi R=1/4$ and
where we choose the constant of integration so that $\tilde{\nu}_1$ is regular at $\gamma=0$.
Unfortunately, we are not able to go any further and find $\tilde{\nu}_2$ from (\ref{eq:rec rln Vk M2xS2}).

In Fig.\ref{fig:Kirchhofff unwrapped cylinder} we compare $V(x,x')$ and $\nu_0+\nu_1\sigma$ with the exact $G_R(x,x')$ as obtained using
an $\ell$-mode sum (see Sec.\ref{sec:Large l}).

In the next section we evaluate (\ref{kirch}) in the PH spacetime and in Sec.\ref{sec:conclusions} we discuss the consequences of the result.


\renewcommand{\thpp}{\theta}

\section{Evaluation of $G_R(x,\xpp)$} \label{sec:Kirchhoff in PH}
\subsection{The basic decomposition.}
We assume henceforth that (\ref{tsep1}) holds. Then, as noted, $G_R(\xp,\xpp)$ and $G_R(x,\xp)$ can be written in Hadamard form:
\be G_R(\xp,\xpp)=[U_1\delta(\sigma_1)+V_1H(-\sigma_1)]H(\tp-\tpp),\label{grxpxpp}\ee
and
\be G_R(x,\xp)=[U_2\delta(\sigma_2)+V_2H(-\sigma_2)]H(t-\tp),\label{grxxp}\ee
where
\begin{eqnarray}
\sigma_1&=&\sigma(\xpp,\xp)=-\frac12\eta_1^2+\frac12\gamma_1^2=-\frac12(\tp-\tpp)^2+\frac12(\yp-\ypp)^2+\frac12\gamma_1^2,\label{sig1form}\\
\sigma_2&=&\sigma(\xp,x)=-\frac12\eta_2^2+\frac12\gamma_2^2=-\frac12(t-\tp)^2+\frac12(y-\yp)^2+\frac12\gamma_2^2 \label{sig2form}
\end{eqnarray}
with an obvious definition of $\eta_1,\gamma_2$, etc. Note that the subindices $1,2$ no longer refer to the different 2-dimensional submanifolds as in Section III.
We also have
\[ U_i=U(\gamma_i)=\left|\frac{\gamma_i}{\sin\gamma_i}\right|^{1/2},\qquad V_i=:\Upsilon(\eta_i,\gamma_i)=\sum_{k=0}^\infty\nu_k(\gamma_i)\sigma_i^k.\]

Our aim is to calculate $G_R(x,\xpp)$ as given by (\ref{kirch}). The following observations are of use in this calculation, and will be used frequently and without necessarily being flagged.

\begin{figure}[h!]
\begin{center}
\includegraphics[width=7cm]
{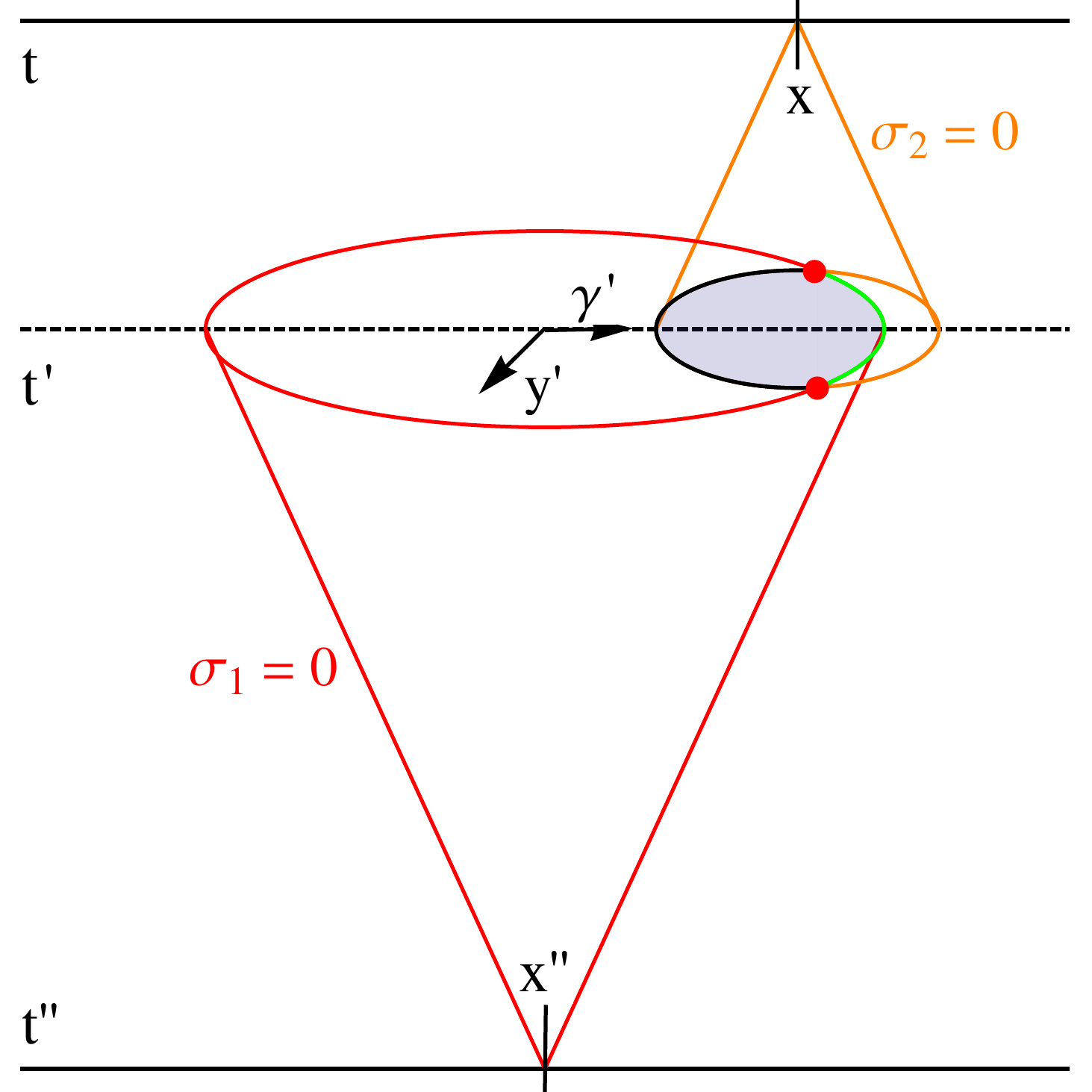}
\end{center}
\caption{
(Colour online).
Diagram representing the $\mathbb{M}_2\times \mathbb{S}^2$ spacetime and illustrating the Kirchhoff integral method.
The time coordinate $T$ runs along the vertical axis, the angular coordinate $\gamma$ (and its negative, $-\gamma$)
runs along the horizontal axis and the $Y$ coordinate
runs perpendicular to the $T$-$\gamma$ axes. Hypersurfaces at three different values of $T$ are represented.
The red curves correspond to the null wavefront $\sigma_1=0$ based at $x''$, while the orange curves correspond to the null
wavefront from $x$. Their overlap on the hypersurface $\{T=t'\}$ is shown with different colours corresponding to the different
contributions to $G_R$ under the Kirchhoff integral method (\ref{Gkdef}): red dots correspond to the `$U\times U$" term,
the green curve (to the right of the dots) and the black curve (to the left of the dots) correspond to different ``$U\times V$" terms and the shaded region corresponds to the ``$V\times V$" terms.
}
\label{fig:Kirchhofff null cones}
\end{figure}

\begin{itemize}
\item[(i)] $G_R(x,\xpp)$ vanishes when $x$ and $\xpp$ are spacelike separated (i.e.\ when there is no causal curve from $\xpp$ to $x$). Thus we are concerned only with points for which
\[ \bar\sigma(\xpp,x)=-\frac12(t-\tpp)^2+\frac12(y-\ypp)^2\leq0.\]
    Equality can be treated as a limiting case. In the (general) case of strict inequality, this indicates that the projections of $x$ and $\xpp$ are timelike separated in $\mt$. Then by a Lorentz transformation in $\mt$, we can arrange $\ypp=y$. This is assumed henceforth.
\item[(ii)] Similarly, by a translation in $\mt$, we can assume that $\ypp=y=0$.
\item[(iii)] By a rotation, we can take $\hpi(\xpp)$ to be at the north pole, so that $\theta^{\prime\prime}=0$. ($\hpi(x)$ is the projection of the spacetime point $x$ onto $\st$.) Notice then that $\gamma_1=\theta^\prime$.
\item[(iv)] By uniqueness, the Kirchhoff representation (\ref{kirch}) of $G_R(x,\xpp)$ is independent of the choice of $\tp\in(\tpp,t)$. It will be convenient to impose an equal time split by setting $\tp=(\tpp+t)/2$ so that
\[ \tp-\tpp=t-\tp = \frac{t-\tpp}{2}=:\frac{\tau}{2}.\]
\item[(v)] The time derivatives $\partial_\tp G_R(\xp,\xpp)$ and $\partial_\tp G_R(x,\xp)$ include terms proportional to $\delta(\tp-\tpp)$ and $\delta(t-\tp)$ respectively. But given the conditions (\ref{tsep1}), these distributions are identically zero for all spacetime points considered. Likewise, $H(\tp-\tpp)=H(t-\tp)\equiv 1$ throughout.
\end{itemize}
Applying these simplifications and using (\ref{grxpxpp}) and (\ref{grxxp}) in (\ref{kirch}), we find
\be
G_R(x,\xpp)=-\frac{1}{4\pi}\sum_{k=1}^{6}G_k,
\label{g1-6}
\ee
where
\be G_k:=\int_\sigtp\cg_k dV^\prime,\quad 1\leq k\leq 6\label{Gkdef}\ee
and
\begin{eqnarray}
\cg_1&:=&\frac{\tau}{2}U_1U_2(\delta(\sigma_1)\delta^\prime(\sigma_2)+\delta^\prime(\sigma_1)\delta(\sigii)),\label{g1def}\\
\cg_2&:=&-\frac{\tau}{2}(U_1V_2+U_2V_1)\delta(\sigi)\delta(\sigii),\label{g2def}\\
\cg_3&:=&\frac{\tau}{2}(U_1V_2\delta^\prime(\sigi)H(-\sigii)+U_2V_1\delta^\prime(\sigii)H(-\sigi)),\label{g3def}\\
\cg_4&:=&U_1\left(\partial_\tp V_2\right)\delta(\sigi)H(-\sigii)-U_2\left(\partial_\tp V_1\right)\delta(\sigii)H(-\sigi),\label{g4def}\\
\cg_5&:=&-\frac{\tau}{2}V_1V_2(\delta(\sigi)H(-\sigii)+\delta(\sigii)H(-\sigi)),\label{g5def}\\
\cg_6&:=&(V_1\partial_\tp V_2-V_2\partial_\tp V_1)H(-\sigi)H(-\sigii).\label{g6def}
\end{eqnarray}
 These are grouped so that (in terms of the coefficients of the distributions) $G_1$ is of the form ``$U\times U$", $G_2-G_4$ are of the form ``$U\times V$" and $G_5$ and $G_6$ are of the form ``$V\times V$". In the following subsections, we evaluate the contribution of these six terms to $G_R(x,\xpp)$.
Fig.\ref{fig:Kirchhofff null cones} illustrates the different contributions to $G_R$ under the Kirchhoff integral method.

Before proceeding with these calculations, we note the following regarding coordinates on the $\st$-component of $\sigtp=\mathbb{R}\times\st$. It is convenient to use the geodesic distances of $\hpi(\xp)$ from $\hpi(x)$ and $\hpi(\xpp)$ as coordinates on this $\st$. So we introduce coordinates $(\alpha,\beta)$ on $\st$ given by
\begin{eqnarray*}
\alpha&:=&\gamma_1(\theta^\prime,\phi^\prime;\theta^{\prime\prime},\phi^{\prime\prime})=\theta^\prime,\\
\beta&:=&\gamma_2(\theta^\prime,\phi^\prime;\thpp,\phi).
\end{eqnarray*}
Using (\ref{gamdef}) - with the appropriate replacements made for $\gamma_2$ - we can show that the invariant volume element on $\st$ is
\be d_2\vp =\omega(\alpha,\beta;\theta)d\alpha d\beta, \quad \omega(\alpha,\beta;\theta) :=  \frac{\sin\alpha\sin\beta}{(\sin^2\alpha-\cos^2\beta+2\cos\thpp\cos\alpha\cos\beta-\cos^2\thpp)^{1/2}},\label{voldef}\ee
and that for a function $Q$ integrable on $\st$,

\begin{eqnarray}
\int_\st Q d_2\vp &=& 2\left\{\int_0^\thpp d\alpha \int_{\thpp-\alpha}^{\thpp+\alpha} Q(\alpha,\beta) \omega(\alpha,\beta)d\beta\right. \nonumber\\
&&+ \int_\thpp^{\pi-\thpp} d\alpha \int_{\alpha-\thpp}^{\alpha+\thpp} Q(\alpha,\beta) \omega(\alpha,\beta)d\beta \nonumber\\
&&+ \left. \int_{\pi-\thpp}^\pi d\alpha \int_{\alpha-\thpp}^{2\pi-\thpp-\alpha} Q(\alpha,\beta) \omega(\alpha,\beta)d\beta\right\}.\label{s2int}
\end{eqnarray}


In fact this decomposition applies only when $0<\thpp<\pi/2$: a similar decomposition applies when $\pi/2<\thpp<\pi$. The calculations below are done for the former case only, but have been checked to apply also to the latter case. We provide a separate (and much shorter) calculation to deal with the case $\gamma(x,\xpp)=\pi$. This yields a significantly different result: see subsection F.

For the upcoming calculations it is important to determine the singularities of the function $\omega(\al,\beta;\theta)$ in (\ref{voldef}).
This function is singular when
\[ \cos\beta = \cos(\al\pm\thpp),\]
that is, when $\beta = \beta_i+2n\pi$ for $1\leq i\leq4$, $n\in\mathbb{Z}$ where
\be \beta_1:= -\al-\thpp,\quad \beta_2:=\thpp-\al,\quad\beta_3:=\al-\thpp,\quad \beta_4:=\al+\thpp.
\label{voldefsing}
\ee
We note that the function $\omega(\al,\beta;\theta)$ diverges near these singularities like $|\beta-\beta_i|^{-1/2}$, $i=1,2,3,4$.

Since the calculations are lengthy, we describe here the main features of the strategy that we follow. We seek to evaluate the triple integrals of (\ref{Gkdef}). The presence of delta functions will sometimes reduce the dimension to two or one. However, all the integrals we encounter share the feature that the integrand becomes infinite, typically at the boundary points of the domain of integration. The strategy we follow is to separate the integrand into singular and non-singular terms. For the latter, we will be able to show that these contribute overall finite terms to $G_R$. For the former, we introduce regularisation procedures that allow us to determine the distributional term corresponding to these singular integrals.

\subsection{The $U\times U$ contribution.}
We have
\be G_1=\int_\sigtp \cg_1 dV^\prime = \int_\mathbb{R}d\yp\int_\st \cg_1 d_2\vp.\label{g11}\ee
We recall that
\[ \delta(f(x))=\sum_{\stackrel{x_n\in\mathbb{R},}{f(x_n)=0}}\frac{\delta(x-x_n)}{|f^\prime(x_n)|}.\]
for any differentiable function $f$.
Then, noting again that $\tp-\tpp>0, t-\tp>0$, we can write
\begin{eqnarray}
\delta(\sigi)&=&\frac{1}{y_1}\{\delta(\yp+y_1)+\delta(\yp-y_1)\}H(\tp-\tpp-\al),\label{delsigi}\\
\delta(\sigii)&=&\frac{1}{y_2}\{\delta(\yp+y_2)+\delta(\yp-y_2)\}H(t-\tp-\beta),\label{delsigii}
\end{eqnarray}
where
\begin{eqnarray}
y_1&=&\sqrt{(\tp-\tpp)^2-\al^2}=\frac12\sqrt{\tau^2-4\al^2},\label{y1} \\
y_2&=&\sqrt{(t-\tp)^2-\beta^2}=\frac12\sqrt{\tau^2-4\beta^2}.\label{y2}
\end{eqnarray}
Then we can carry out the $\yp$ integration in (\ref{g11}) - where we note that the integrands of the two terms concerned are both even functions of $\yp$. We find
\begin{eqnarray}
\frac{2}{\tau}G_1 &=&\int_\mathbb{R}d\yp\int_\st U(\alpha)U(\beta)[\delta(\sigi)\delta^\prime(\sigii)+\delta(\sigi)\delta^\prime(\sigii)]d_2\vp\nonumber \\
&=&2\int_\st U(\al)U(\beta)\left[\frac{\delta^\prime(\sigii)|_{\yp=y_1}}{y_1}H(\tau-2\al)+\frac{\delta^\prime(\sigi)|_{\yp=y_2}}{y_2}H(\tau-2\beta)\right]d_2\vp\nonumber\\
&=&4\int_\st \frac{U(\al)U(\beta)}{\al\beta}\left[\frac{H(\tau-2\al)}{\sqrt{\tau^2-4\al^2}}-\frac{H(\tau-2\beta)}{\sqrt{\tau^2-4\beta^2}}\right]\partial_\beta\{\delta(\beta-\al)\}d_2\vp\nonumber\\
&=&-4\int_\st\partial_\beta\left\{\omega(\al,\beta;\theta)\frac{U(\al)U(\beta)}{\al\beta}\left[\frac{H(\tau-2\al)}{\sqrt{\tau^2-4\al^2}}-\frac{H(\tau-2\beta)}{\sqrt{\tau^2-4\beta^2}}\right]\right\}\delta(\beta-\al)d\al d\beta\nonumber\\
&=&4\int_\st \underbrace{F(\al,\beta;\theta)\partial_\beta\left\{\frac{H(\tau-2\beta)}{\sqrt{\tau^2-4\beta^2}}\right\}}_{J(\al,\beta;\theta,\tau)}\delta(\beta-\al)d\al d\beta,
\label{g12}
\end{eqnarray}
where
\be F(\al,\beta;\theta):=\omega(\al,\beta;\theta)\frac{U(\al)U(\beta)}{\al\beta}.\label{hdef}\ee

Applying the decomposition (\ref{s2int}), we can carry out the $\beta-$integration to obtain
\be G_1=4\tau\int_{\thpp/2}^{\pi-\thpp/2}J(\al,\al;\theta,\tau)d\alpha.\label{g13}\ee

We know that $G_R(\xpp,x)$ - not necessarily as generated by (\ref{kirch}) - is a distribution, and so we should not expect the integral (\ref{g13}) to converge to a function. However we do expect it to be a distribution. This is indeed the case, and to identify this distribution, we introduce the following regularization procedure.
From (\ref{voldefsing}) we see that the integrand is singular at the limits of integration, therefore we let
\be G_1=\lim_{\epsilon \to 0^+}G_1^{(\epsilon)},\label{g1ep}\ee
where
\[ G_1^{(\ep)} := 4\tau\int_{\thpp/2+\ep}^{\pi-\thpp/2-\ep}J(\al,\al;\theta,\tau)d\al.\]
Integrating by parts yields
\begin{eqnarray}
\frac{1}{4\tau}G_1^{(\ep)}
&=& -J_1+J_2-I \label{IandJ}
\end{eqnarray}

where

\begin{eqnarray}
J_1 &=& \frac{H(\tau-\thpp-2\ep)}{\sqrt{\tau^2-(\thpp+2\ep)^2}}\frac{1}{\thpp\sqrt{\sin\thpp}}\left[\frac{1}{\sqrt{\ep}}+O(\ep^{1/2})\right],\label{J1def}\\
J_2&=&\frac{H(\tau-2\pi+\thpp+2\ep)}{\sqrt{\tau^2-(2\pi-\thpp-2\ep)^2}}\frac{1}{(2\pi-\thpp)\sqrt{\sin\thpp}}\left[\frac{1}{\sqrt{\ep}}+O(\ep^{1/2})\right],\label{J2def}\\
I&=&\int_{\thpp/2+\ep}^{\pi-\thpp/2-\ep}\frac{Q(\al,\theta)}{\sqrt{\tau^2-4\al^2}}H(\tau-2\al)d\al\label{Idef}
\end{eqnarray}

and
\be Q(\al,\theta):=\frac{d}{d\al}\{F(\al,\al;\theta)\},\quad F(\al,\al;\theta) = \frac{1}{\sqrt{2}\sin(\theta/2)}\frac{\sin\al}{\al(\cos\theta-\cos2\al)^{1/2}}.\label{qdef}\ee
We note that for each fixed $\thpp$, $Q$ is analytic on $(\frac{\thpp}{2},\pi-\frac{\thpp}{2})$, but is singular at the endpoints of this interval.
Also, in (\ref{J1def}) and (\ref{J2def}), $O$ is the usual Lifschitz big-Oh, and the asymptotic relation holds in the limit $\ep\to0^+$.

Removing the step function from the integrand of $I$ yields
\be
I = H(\tau-\thpp-2\ep)H(-\tau+2\pi-\theta-2\ep)\int_{\theta/2+\ep}^{\tau/2} \frac{Q(\al,\theta)}{\sqrt{\tau^2-4\al^2}}d\al
+H(\tau-2\pi+\theta+2\ep)\int_{\theta/2+\ep}^{\pi-\theta/2-\ep} \frac{Q(\al,\theta)}{\sqrt{\tau^2-4\al^2}}d\al.
\label{Inostep1}
\ee



It is appropriate to pause at this juncture to consider the result that we expect to obtain for $G_1$. Given that $G_1$ is constructed from the singular parts of $G_R(\xp,\xpp)$ (field) and $G_R(x,\xp)$ (propagator), we expect it to correspond to the ``most" singular part of $G_R(x,\xpp)$. On general grounds, we expect that the spacetime location of the singularity in $G_1$ is a subset of the future null cone of $\xpp$. That is, we expect that $G_1(x,\xpp)$ is singular at points $x$ which are connected to $\xpp$ by a null geodesic. Using the notation above, and recalling that $\theta^{\prime\prime}=0$, these are points for which
\[ \tau = (-1)^n\thpp+2n\pi,\quad n\geq 0.\]
Here, $n$ counts the number of caustic points through which the geodesic has passed: for $n=1$, the geodesic has passed through the caustic $x$ with $\hpi(x)$ antipodal to $\hpi(\xpp)$; for $n=2$, it has passed through this caustic and that with $\hpi(x)=\hpi(\xpp)$, etc. From the form of $G_1^{(\ep)}$ above, and the expression (\ref{Inostep1}) for $I$, it is clear that these are `points of interest' in our investigation.
We consider the case of most interest, which corresponds to $\tau\in(\pi,2\pi)$. In this case, $x$ lies outside the maximal normal neighbourhood of $\xpp$, and the null geodesic spray from $\xpp$ has formed exactly one caustic. Then $n=1$, and we expect to see that $G_R(x,\xpp)$ is singular when, and only when, $\tau=2\pi-\theta$.

We recall that our aim is to calculate the singular part of $G_R(x,\xpp)$. In this calculation, we will frequently encounter terms $A$ that are finite for all values of the spacetime variables (perhaps of the region of spacetime presently under consideration, e.g. $\pi<\tau<2\pi$), and for all sufficiently small values of $\ep>0$, and that remain finite in the limit $\ep\to0^+$. Such terms do not, by definition, contribute to the singular part of $G_R$. We use the following notation to characterise quantities which possess these features:
\be A = \co(1).\label{codef}\ee

Using (\ref{Inostep1}) and noting that $\tau>\pi>\theta$, and introducing (for ease of notation) $a:=2\al$, we have
\be
I = H(-\tau+2\pi-\theta-2\ep)I_1+H(\tau-2\pi+\theta+2\ep)I_2
\label{Inostep2}
\ee
where
\be
I_1:= \int_{\theta+2\ep}^{\tau} \frac{R(a,\theta)}{\sqrt{\tau^2-a^2}}da,\qquad I_2:=\int_{\theta+2\ep}^{2\pi-\theta-2\ep} \frac{R(a,\theta)}{\sqrt{\tau^2-a^2}}da\label{I12def}
\ee

and

\be R(a,\theta):=\frac12Q\left(\frac{a}{2},\theta\right) = \frac{1}{\sqrt{2}\sin(\theta/2)}\left[
\frac{(a\cos(a/2)-2\sin(a/2))(\cos\theta-\cos a)-a\sin(a/2)\sin a}{a^2(\cos\theta-\cos a)^{3/2}}\right].
\label{Radef}
\ee

Notice then that $R(a,\theta)$ is singular at the points $a=\theta$ and $a=2\pi-\theta$. These correspond to the lower and upper limits of $I_2$, and to the lower limit of $I_1$, when $\ep\to 0^+$. Further, $a=2\pi-\theta$ also corresponds to the upper limit of $I_1$ in the limiting case $\tau = 2\pi-\theta$, $\ep\to0^+$. So in order to capture the dominant contribution to $I$, we need to isolate the singular behaviour in $R(a,\theta)$, leaving a finite remainder that will contribute an overall $\co(1)$ term. In fact we proceed by isolating the singular behaviour of $R(a,\theta)/\sqrt{\tau+a}$. This greatly simplifies the resulting integrals. Notice that the factor $\frac{1}{\sqrt{\tau+a}}$ is analytic on $[\theta,2\pi-\theta]$ and so does not introduce any additional singular behaviour.

We define
\begin{eqnarray}
R_1(\theta,\tau) &:=& \lim_{a\to \theta^+} (a-\theta)^{3/2}\frac{R(a,\theta)}{\sqrt{\tau+a}},\label{r1def}\\
R_2(\theta,\tau) &:=& \lim_{a\to\theta^+} (a-\theta)^{1/2}\left[\frac{R(a,\theta)}{\sqrt{\tau+a}}-\frac{R_1(\theta,\tau)}{(a-\theta)^{3/2}}\right],\label{r2def}\\
R_3(\theta,\tau) &:=& \lim_{a\to(2\pi-\theta)^-} (2\pi-\theta-a)^{3/2}\frac{R(a,\theta)}{\sqrt{\tau+a}},\label{r3def}\\
R_4(\theta,\tau) &:=& \lim_{a\to(2\pi-\theta)^-} (2\pi-\theta-a)^{1/2}\left[\frac{R(a,\theta)}{\sqrt{\tau+a}}-\frac{R_3(\theta,\tau)}{(2\pi-\theta-a)^{3/2}}\right]\label{r4def}.
\end{eqnarray}
Of these, the following values are required:
\begin{eqnarray}
R_1(\theta,\tau) &=&-\frac{1}{\sqrt{2}\theta\sqrt{\sin\theta}\sqrt{\tau+\theta}},\label{r1val}\\
R_3(\theta,\tau) &=& \frac{1}{\sqrt{2}(2\pi-\theta)\sqrt{\sin\theta}\sqrt{\tau+2\pi-\theta}},\label{r3val}\\
R_4(\theta,\tau) &=& -\frac{(\tau+2\pi-\theta)(2\pi-\theta)(2+\cos\theta)+2(2\tau+2\pi-\theta)\sin\theta}{4\sqrt{2}(2\pi-\theta)^2(\tau+2\pi-\theta)^{3/2}(\sin\theta)^{3/2}}.\label{r4val}
\end{eqnarray}
The $R_i$ are the coefficients of the divergent parts of $R/\sqrt{\tau+a}$ at the different endpoints. It follows that we can write
\be \frac{R(a,\theta)}{\sqrt{\tau+a}} = \frac{R_1(\theta,\tau)}{(a-\theta)^{3/2}}+\frac{R_2(\theta,\tau)}{(a-\theta)^{1/2}}+\frac{R_3(\theta,\tau)}{(2\pi-\theta-a)^{3/2}}+\frac{R_4(\theta,\tau)}{(2\pi-\theta-a)^{1/2}}+\tilde{R}(a,\theta,\tau),
\label{tilrdef}\ee
where for each $\tau\in(\pi,2\pi)$ and each $\theta\in(0,\pi/2)$, $\tilde{R}(a,\theta,\tau)$ is a continuous function of $a$ on the interval $[\theta,2\pi-\theta]$. It follows that
\be \int_{\theta+2\ep}^{\tau} \frac{\tilde{R}(a,\theta,\tau)}{\sqrt{\tau-a}}da=\int_{\theta+2\ep}^{2\pi-\theta-2\ep} \frac{\tilde{R}(a,\theta,\tau)}{\sqrt{\tau-a}}da = \co(1).\label{rtilint}\ee
That is, $\tilde{R}$ makes only an $\co(1)$ contribution to $I$. The following antiderivatives are then needed (constants of integration are omitted):
\begin{eqnarray}
\int\frac{da}{\sqrt{\tau-a}(a-\theta)^{3/2}} &=& \frac{2\sqrt{\tau-a}}{(\theta-\tau)\sqrt{a-\theta}},\qquad \int\frac{da}{\sqrt{\tau-a}(a-\theta)^{1/2}}=2\arctan\left(\frac{\sqrt{a-\theta}}{\sqrt{\tau-a}}\right),\label{ad1}\\
\int\frac{da}{\sqrt{\tau-a}(2\pi-\theta-a)^{3/2}}&=&\frac{2\sqrt{\tau-a}}{(\tau-2\pi+\theta)\sqrt{2\pi-\theta-a}},\qquad
\int\frac{da}{\sqrt{\tau-a}(2\pi-\theta-a)^{1/2}}=-2\ln\left(\sqrt{\tau-a}+\sqrt{2\pi-\theta-a}\right).
\label{ad2}
\end{eqnarray}
Applying these (and using (\ref{rtilint})), we see that all terms proportional to $R_2$ are $\co(1)$, and we obtain
\begin{eqnarray} I_1 &=& -2R_1\frac{\sqrt{\tau-\theta-2\ep}}{(\theta-\tau)\sqrt{2\ep}}-2R_3\frac{\sqrt{\tau-\theta-2\ep}}{(\tau-2\pi+\theta)\sqrt{2\pi-2\theta-2\ep}}
\nn\\
&& - 2R_4\left\{\ln(\sqrt{2\pi-\theta-\tau})-\ln(\sqrt{\tau-\theta-2\ep}+\sqrt{2\pi-2\theta-2\ep})\right\}+\co(1)
\label{I1form}
\end{eqnarray}
and
\begin{eqnarray}
I_2 &=& -2R_1\frac{\sqrt{\tau-\theta-2\ep}}{(\theta-\tau)\sqrt{2\ep}}
+\frac{2R_3}{(\tau-2\pi+\theta)}\left\{\frac{\sqrt{\tau-2\pi+\theta+2\ep}}{\sqrt{2\ep}}-\frac{\sqrt{\tau-\theta-2\ep}}{\sqrt{2\pi-2\theta-2\ep}}\right\}
\nn\\
&& -2R_4\left\{\ln(\sqrt{\tau-2\pi+\theta+2\ep}+\sqrt{2\ep})-\ln(\sqrt{\tau-\theta-2\ep}+\sqrt{2\pi-2\theta-2\ep})\right\}+\co(1)
\label{I2form}
\end{eqnarray}

 Now we collect terms from (\ref{J1def}), (\ref{J2def}), (\ref{I1form}) and (\ref{I2form}) to evaluate $G_1^{(\ep)}$ via (\ref{IandJ}) and (\ref{Inostep2}).

We note first that the $R_1$ term in $I$ cancels to $O(\ep)$ with $J_1$ (using $\tau>\theta$ in $J_1$), and so there is only an $\co(1)$ contribution to $G_1$ from these terms. Noting that the $R_1$ contribution to $I$ arises by formally setting $R_2=R_3=R_4=0$, we can write
\be
\lim_{\ep\to0^+} (J_1 + I|_{R_2=R_3=R_4=0}) = \co(1).\label{R1lim}\ee

The contributions from $J_2$ and the $R_3$ terms in $I$ arise by formally setting $R_1=R_2=R_4=0$ in $I$. These yield
\begin{eqnarray}
J_2-I|_{R_1=R_2=R_4=0}&=&\frac{1}{(2\pi-\theta)\sqrt{\sin\theta}\sqrt{\tau+2\pi-\theta}}\times\left\{H(-\tau+2\pi-\theta-2\ep)\left[\frac{\sqrt{\tau-\theta-2\ep}}{\sqrt{\pi-\theta-\ep}}\frac{1}{\tau-2\pi+\theta}\right]
\right.\nn\\
&&+\left.H(\tau-2\pi+\theta+2\ep)\left[\left(\frac{1}{\sqrt{\tau-2\pi+\theta+2\ep}}-\frac{\sqrt{\tau-2\pi+\theta+2\ep}}{\tau-2\pi+\theta}\right)\frac{1}{\sqrt{\ep}}
+\frac{\sqrt{\tau-\theta-2\ep}}{\sqrt{\pi-\theta-\ep}}\frac{1}{\tau-2\pi+\theta}\right]\right\}\nn\\
&&+\co(1)
\label{R3terms}
\end{eqnarray}

To evaluate the terms in braces here it is convenient to introduce $z:=\tau-2\pi+\theta$. Note then
\begin{eqnarray*} \frac{\sqrt{\tau-\theta-2\ep}}{\sqrt{\pi-\theta-\ep}}\frac{1}{\tau-2\pi+\theta} &=& \frac{\sqrt{2\pi-2\theta-2\ep+z}}{\sqrt{\pi-\theta-\ep}}\frac{1}{z}\\
&=&\frac{\sqrt{2}}{z}+\co(1),
\end{eqnarray*}
and so we can write
\be J_2-I|_{R_1=R_2=R_4=0}=\frac{1}{(2\pi-\theta)\sqrt{\sin\theta}\sqrt{\tau+2\pi-\theta}}\Delta_1^{(\ep)}(z)+\co(1)\ee
where
\be
\Delta_1^{(\ep)}(z):=\frac{\sqrt{2}}{z}H(-z-2\ep)+\frac{\sqrt{2}}{z}\left(1-\frac{\sqrt{2\ep}}{\sqrt{z+2\ep}}\right)H(z+2\ep).\label{del1def}
\ee
In Appendix A.1, we derive the distributional identity
\be \lim_{\ep\to0^+} \Delta_1^{(\ep)}(z) = \sqrt{2}PV\left(\frac1z\right)\label{del1lim}\ee
where $PV$ is the principal value distribution, defined by the following action on a test function $\phi$:
\[ <PV\left(\frac1z\right),\phi>=\lim_{\ep\to 0^+} \int_{\mathbb{R}\setminus[-\ep,\ep]}\frac{\phi(z)}{z}dz.\]
Thus
\be \lim_{\ep\to0^+}(J_2-I|_{R_1=R_2=R_4=0})=\frac{\sqrt{2}}{(2\pi-\theta)\sqrt{\sin\theta}\sqrt{\tau+2\pi-\theta}}PV\left(\frac{1}{\tau-2\pi+\theta}\right)+\co(1)
\label{R3lim}\ee

Finally, we consider the $R_4$ contributions to $G_1^{(\ep)}$. These take the form of logarithms. We note that in (\ref{I1form}) and (\ref{I2form}), due to the range of values of $\tau$ and $\theta$ under consideration, we have
\[ \ln|\sqrt{\tau-\theta-2\ep}+\sqrt{2\pi-2\theta-2\ep}|=\co(1).\]
We also have the distributional identity (see Appendix A.2)
\be \lim_{\ep\to0^+} \ln(\sqrt{z+2\ep}+\sqrt{2\ep})H(z+2\ep)=\ln(\sqrt{z})H(z).\label{heplim}\ee
It follows that
\begin{eqnarray}\lim_{\ep\to0^+} I|_{R_1=R_2=R_3=0} &=& -R_4\ln|\tau-2\pi+\theta|(H(-\tau+2\pi-\theta)+H(\tau-2\pi+\theta))\nn\\
&=& -R_4\ln|\tau-2\pi+\theta|.
\label{R4lim}\end{eqnarray}

Collecting the results of (\ref{R1lim}), (\ref{R3lim}) and (\ref{R4lim}) completes the calculation of $G_1$:
\begin{eqnarray} G_1=\lim_{\ep\to0^+} G_1^{(\ep)} &=& \frac{4\sqrt{2}\tau}{(2\pi-\theta)\sqrt{\sin\theta}\sqrt{\tau+2\pi-\theta}}\times\nn\\
&&\left\{PV\left(\frac{1}{\tau-2\pi+\theta}\right)-\frac{(\tau+2\pi-\theta)(2\pi-\theta)(2+\cos\theta)+2(2\tau+2\pi-\theta)\sin\theta}{8(2\pi-\theta)(\tau+2\pi-\theta)\sin\theta}\ln|\tau-2\pi+\theta|\right\}+\co(1)\nn\\
&=&\frac{4}{\sqrt{\sin\theta}\sqrt{2\pi-\theta}}\left\{PV\left(\frac{1}{\tau-2\pi+\theta}\right)-\frac{(2\pi-\theta)(2+\cos\theta)+3\sin\theta}{8(2\pi-\theta)\sin\theta}\ln|\tau-2\pi+\theta|\right\}+\co(1).\label{G1-result}
\end{eqnarray}
In the last line here, we have used $f(x)PV(\frac1x)=f(0)PV(\frac1x)+\co(1)$ and similar with $PV$ replaced by $\ln$, for a differentiable function $f$. That is, we set $\tau=2\pi-\theta$ in the coefficients of $PV$ and $\ln$.

\subsection{The $V\times V$ contribution.}

Two terms contribute to $G_R(x,\xpp)$ under this heading, corresponding to $G_5$ and $G_6$. We show here that
\be G_5=\co(1),\qquad G_6=\co(1) \ee
for all $\tau\in(0,2\pi)$.

We consider $G_6$ first. Using the functional forms of $\sigma_i$ and $V_i$ ($i=1,2$) given in (\ref{sig1form}), (\ref{sig2form}) and (\ref{vdef}), we see that
\be \partial_\tp V_1 =-\frac{\tau}{2\yp}\partial_\yp V_1,\qquad\partial_\tp V_2=\frac{\tau}{2\yp}\partial_\yp V_2.\label{vpartials}\ee
Then combining (\ref{Gkdef}) and (\ref{g6def}) gives
\[ G_6 = -\frac{\tau}{2}\int_\sigtp dV^\prime\left(\frac{V_1}{\yp}\partial_\yp V_2+\frac{V_2}{\yp}\partial_\yp V_1\right)H(-\sigi)H(-\sigii).\]
Noting that each $V_i$ is an even smooth function of $\yp$, we see that this is the integral of a smooth function over the compact set $\{\xp\in\sigtp: \sigi(x,\xp)\leq0 \hbox{ and } \sigii(\xp,\xpp)\leq0\}$. Hence
\be G_6=\co(1).\label{G6-result}\ee

From (\ref{Gkdef}) and (\ref{g5def}), and using (\ref{delsigi}) and (\ref{delsigii}), we have
\bes G_5&=&-\tau\int_\st d_2\vp\int_0^\infty d\yp\left[\frac{V_1V_2}{y_1}\delta(\yp-y_1)H(\tau-2\alpha)H(-\sigii)+\frac{V_1V_2}{y_2}\delta(\yp-y_2)H(\tau-2\beta)H(-\sigi)\right]\\
&=&-2\tau\int_\st d_2\vp \left[\nu_0(\al)V_2|_{\yp=y_1}\frac{H(\tau-2\al)}{\sqrt{\tau^2-4\al^2}}H(\al-\beta)+\nu_0(\beta)V_1|_{\yp=y_2}\frac{H(\tau-2\beta)}{\sqrt{\tau^2-4\beta^2}}H(\beta-\al)\right].
\ees
where $y_1$ and $y_2$ are given in (\ref{y1}) and (\ref{y2}).
The terms $\nu_0(\al)V_2|_{\yp=y_1}$ and $\nu_0(\beta)V_1|_{\yp=y_2}$ are analytic on $\st$, and so both integrals are finite: the integrand diverges at $\tau=2\al$ and at $\tau=2\beta$, but only at the rate $(\tau-2\al)^{-1/2}$ (respectively $(\tau-2\beta)^{-1/2}$), yielding an overall integrable term. Hence we can state that
\be G_5=\co(1).\label{G5-result}\ee

\subsection{The $U\times V$ contribution.}

The terms $G_2$, $G_3$ and $G_4$ form the $U\times V$ contribution.
From (\ref{sig1form}) and (\ref{sig2form}), we see that
\[\delta^\prime(\sigma_i) = \frac{1}{\yp}\partial_\yp\left\{\delta(\sigma_i)\right\}.\] Using this and (\ref{vpartials}), we can combine (\ref{g3def}) and (\ref{g4def}) to obtain
\be
G_3+G_4=\frac{\tau}{2}\int_\st d_2\vp\int_\re d\yp\left[\frac{U(\alpha)}{\yp}H(-\sigii)\partial_\yp\left\{V_2\delta(\sigi)\right\} + 1 \leftrightarrow 2 \right], \label{g34first}
\ee
where here and below, $1\leftrightarrow2$ indicates duplication of the preceding expression with the replacements $1\leftrightarrow2$ and $\alpha\leftrightarrow\beta$. The pole at the origin shows that, as above, the integral must be regularized. We write
$G_2+G_3+G_4=\lim_{\ep\to0^+}G_{2,3,4}^{(\ep)}$, where
\be
G_{3,4}^{(\ep)}:=\frac{\tau}{2}\int_\st d_2\vp\left\{\int_{-\infty}^{-\ep}+\int_\ep^\infty\right\} d\yp\left[\frac{U(\alpha)}{\yp}H(-\sigii)\partial_\yp\left\{V_2\delta(\sigi)\right\} + 1 \leftrightarrow 2 \right]. \label{g34ep}
\ee

To proceed, we integrate by parts. This introduces three types of term: boundary terms corresponding to $\yp=\pm\ep$ (these are collected in $K_1$ below), terms arising from the derivative $\partial_\yp$ acting on $1/\yp$ (these are collected in $K_2$ below) and terms arising from the derivative acting on $H(-\sigma_i)$. These last yield terms with a product of delta functions in the integrand. These terms cancel identically with the $G_2^{(\ep)}$ contribution (with the obvious definition of $G_2^{(\ep)}$ - the integral (\ref{Gkdef}) with $k=2$ and a principal value truncation on the $\yp$ axis). Thus we can write
\[ G_2+G_3+G_4 = \lim_{\ep\to0^+}G_{2,3,4}^{(\ep)}\]
where
\be G_{2,3,4}^{(\ep)} = K_1+K_2,\label{g34epk}\ee
with
\begin{eqnarray}
K_1 &:=& -\tau\int_\st d_2\vp\left[U(\alpha)\left.\frac{V_2}{\ep}\right|_{\yp=\ep}\delta\left(-\frac{\tau^2}{8}+\frac{\ep^2}{2}+\frac{\alpha^2}{2}\right)H\left(\frac{\tau^2}{8}-\frac{\ep^2}{2}-\frac{\beta^2}{2}\right)
+ 1\leftrightarrow 2\right]\nonumber\\
&=&-\tau\int_\st d_2\vp\left[U(\alpha)\left.\frac{V_2}{\ep}\right|_{\yp=\ep}\delta\left(-\frac{\tau^2}{8}+\frac{\ep^2}{2}+\frac{\alpha^2}{2}\right)H(\al-\beta)
+ 1\leftrightarrow 2\right]\nonumber\\
&=&-2\tau\int_\st d_2\vp U(\alpha)\left.\frac{V_2}{\ep}\right|_{\yp=\ep}\delta\left(-\frac{\tau^2}{8}+\frac{\ep^2}{2}+\frac{\alpha^2}{2}\right)H(\al-\beta),\label{K1def}
\end{eqnarray}
and
\begin{eqnarray}
K_2 &:=& \tau\int_\st d_2\vp\int_\ep^\infty d\yp\left[ \frac{U(\alpha)V_2}{\yp^2}\delta(\sigi)H(-\sigii) + 1\leftrightarrow 2\right] \nonumber \\
&=&\tau\int_\st d_2\vp\left[\frac{U(\alpha)}{y_1^3}V_2|_{\yp=y_1}H(\tau-2\alep)H(\alpha-\beta) + 1\leftrightarrow 2\right]\nonumber\\
&=&2\tau\int_\st d_2\vp \frac{U(\alpha)}{y_1^3}V_2|_{\yp=y_1}H(\tau-2\alep)H(\alpha-\beta),
\label{K2def}
\end{eqnarray}
with
\be \alep:=\sqrt{\al^2+\ep^2}.\label{alepdef}\ee

The first expressions for each of $K_1, K_2$ follow by noting that the integrand is an even function of $\yp$. The second expression for $K_1$ relies on the fact that there is a contribution only at the zero of the argument of the delta function, and that $\al,\beta$ are both positive so that $H(\al^2-\beta^2)=H(\al-\beta)$. The final expressions arise because the $1\leftrightarrow2$ contributions are equal. (This last fact is not immediately obvious, but follows from a decomposition equivalent to (\ref{s2int}) in which the order of integration with respect to $\al,\beta$ is reversed.)

It is useful to consider the general form of the integrals $K_1, K_2$: using the decomposition (\ref{s2int}), it can be shown that for any function $f$ that is integrable on $\st$,
\begin{eqnarray}
\int_\st d_2\vp f(\al,\beta)H(\al-\beta) &=& 2\int_{\thpp/2}^\thpp d\al \int_{\thpp-\al}^\al f(\al,\beta)\omega(\al,\beta)d\beta \nonumber \\
&&+ 2\int_{\thpp}^{\pi-\thpp/2} d\al \int_{\al-\thpp}^\al f(\al,\beta)\omega(\al,\beta)d\beta \nonumber \\
&&+ 2\int_{\pi-\thpp/2}^\pi d\al \int_{\al-\thpp}^{2\pi-\thpp-\al} f(\al,\beta)\omega(\al,\beta)d\beta.
\label{theta-int-s2}
\end{eqnarray}

We will then naturally encounter these terms:
\begin{eqnarray}
L_1(\al) &:=& \int_{\thpp-\al}^\al \omega(\al,\beta)\nu(\al,\beta) d\beta,\label{L1def}\\
L_2(\al) &:=& \int_{\al-\thpp}^\al \omega(\al,\beta)\nu(\al,\beta) d\beta,\label{L2def}\\
L_3(\al) &:=& \int_{\al-\thpp}^{2\pi-\thpp-\al} \omega(\al,\beta)\nu(\al,\beta) d\beta,\label{L3def}
\end{eqnarray}
where
\be \nu(\al,\beta) := \sum_{n=0}^\infty \nu_n(\beta)\left(\frac{\beta^2-\al^2}{2}\right)^n. \label{nudef}\ee

The term $K_1$ corresponds to an integral of the form (\ref{theta-int-s2}) with
\begin{eqnarray}
f(\al,\beta)&=& -2\tau U(\al)\left.\frac{V_2}{\ep}\right|_{\yp=\ep}\delta\left(-\frac{\tau^2}{8}+\frac{\ep^2}{2}+\frac{\alpha^2}{2}\right)\nonumber\\
&=&-\frac{4\tau}{\ep\tauep}U({\tauep/2})\delta(\al-{\tauep/2})V_2|_{\yp=\ep,\al=\tauep/2}\nonumber\\
&=&-\frac{4\tau}{\ep\tauep}U({\tauep/2})\delta(\al-{\tauep/2})\nu(\tauep/2,\beta),
\end{eqnarray}
where
\be\tauep := \sqrt{\tau^2-4\ep^2}.\label{tauepdef}\ee

Then using (\ref{theta-int-s2}) and the definitions (\ref{L1def})-(\ref{L3def}), we find
\begin{eqnarray}
K_1 &=& - \frac{8\tau}{\ep\tauep}U({\tauep/2})\left\{ H(\tauep-\thpp)H(\thpp-{\tauep/2})L_1({\tauep/2})\right.\nonumber\\
&&+ H({\tauep/2}-\thpp)H(2\pi-\thpp-\tauep)L_2({\tauep/2})\nonumber\\
&&+ \left.H(\tauep-2\pi+\thpp)H(2\pi-\tauep)L_3({\tauep/2})\right\}.\label{K1-in-L}
\end{eqnarray}

Likewise,
\begin{eqnarray}
K_2 &=& 32\tau\left\{ \int_{\thpp/2}^\thpp \frac{U(\al)}{(\tau^2-4\al^2)^{3/2}}L_1(\al)H(\tau-2\alep)d\al\right.\nonumber\\
&&+\int_{\thpp}^{\pi-\thpp/2} \frac{U(\al)}{(\tau^2-4\al^2)^{3/2}}L_2(\al)H(\tau-2\alep)d\al\nonumber\\
&&+\left.\int_{\pi-\thpp/2}^\pi \frac{U(\al)}{(\tau^2-4\al^2)^{3/2}}L_3(\al)H(\tau-2\alep)d\al\right\}\label{K2-in-L}
\end{eqnarray}
The terms $H(\tau-2\alep)$ here arise naturally, but for the next steps we note that $H(\tau-2\alep)=H({\tauep/2}-\al)$. This follows from the definitions (\ref{alepdef}), (\ref{tauepdef}). The step functions can be extracted from the integrals to obtain
\begin{eqnarray}
K_2 &=& 32\tau\left\{H(\tauep-\thpp)H(\thpp-{\tauep/2})\int_{\thpp/2}^{\tauep/2}\frac{U(\al)}{(\tau^2-4\al^2)^{3/2}}L_1(\al)d\al\right.\nonumber\\
&&+ H({\tauep/2}-\thpp)\int_{\thpp/2}^\thpp\frac{U(\al)}{(\tau^2-4\al^2)^{3/2}}L_1(\al)d\al\nonumber\\
&&+ H({\tauep/2}-\thpp)H(2\pi-\thpp-\tauep)\int_{\thpp}^{\tauep/2}\frac{U(\al)}{(\tau^2-4\al^2)^{3/2}}L_2(\al)d\al\nonumber\\
&&+ H(\tauep-2\pi+\thpp)\int_{\thpp}^{\pi-\thpp/2}\frac{U(\al)}{(\tau^2-4\al^2)^{3/2}}L_2(\al)d\al\nonumber\\
&&+ H(\tauep-2\pi+\thpp)H(2\pi-\tauep)\int_{\pi-\thpp/2}^{\tauep/2}\frac{U(\al)}{(\tau^2-4\al^2)^{3/2}}L_3(\al)d\al\nonumber\\
&&+ \left.H(\tauep-2\pi)\int_{\pi-\thpp/2}^{\pi}\frac{U(\al)}{(\tau^2-4\al^2)^{3/2}}L_3(\al)d\al\right\}.\label{K2-in-L-full}
\end{eqnarray}

We consider from now on the the case of most interest, when $\pi<\tau<2\pi$, so that the point $x$ lies outside the maximal normal neighbourhood of $\xpp$, after the formation of the first caustic but prior to the formation of the second caustic. Hence for sufficiently small $\ep$, we have $\pi<\tauep<2\pi$ and since $0<\thpp<\pi/2$, we also have $\tauep>2\thpp>\thpp$. Then some terms of (\ref{K1-in-L}) and (\ref{K2-in-L-full}) are identically zero and others simplify, leaving

\begin{eqnarray}
K_1 &=& K_{1(ii)}+K_{1(iv)}\label{K1split}
\end{eqnarray}

where

\begin{eqnarray}
K_{1(ii)} &:=& - \frac{8\tau}{\ep\tauep}U({\tauep/2})H(2\pi-\thpp-\tauep)L_2({\tauep/2}),\label{K12def}\\
K_{1(iv)} &:=&- \frac{8\tau}{\ep\tauep}U({\tauep/2})H(\tauep-2\pi+\thpp)L_3({\tauep/2})\label{K14def}
\end{eqnarray}

and

\begin{eqnarray}
K_2 &=& K_{2(i)}+K_{2(ii)}+K_{2(iii)}+K_{2(iv)}\label{K2split}
\end{eqnarray}

with

\begin{eqnarray}
K_{2(i)} &:=& 32\tau\int_{\thpp/2}^\thpp\frac{U(\al)}{(\tau^2-4\al^2)^{3/2}}L_1(\al)d\al,\label{K21def}\\
K_{2(ii)} &:=&32\tau H(2\pi-\thpp-\tauep)\int_{\thpp}^{\tauep/2}\frac{U(\al)}{(\tau^2-4\al^2)^{3/2}}L_2(\al)d\al,\label{K22def}\\
K_{2(iii)} &:=& 32\tau H(\tauep-2\pi+\thpp)\int_{\thpp}^{\pi-\thpp/2}\frac{U(\al)}{(\tau^2-4\al^2)^{3/2}}L_2(\al)d\al,\label{K23def}\\
K_{2(iv)} &:= &32\tau H(\tauep-2\pi+\thpp)\int_{\pi-\thpp/2}^{\tauep/2}\frac{U(\al)}{(\tau^2-4\al^2)^{3/2}}L_3(\al)d\al.\label{K24def}
\end{eqnarray}


The nature of the integrals contributing to $K_1$ and $K_2$ depends on whether or not
the singular points (\ref{voldefsing}) of the function $\omega(\al,\beta)$ occur in the domain of the integrals.
  In some of the cases below, we see that the singularities occur only in certain limits. It is nevertheless crucial to take account of these, as the overall singular contributions to $G_R(x,\xpp)$ will include contributions from these limiting singular values. As with the calculation of the $U\times U$ contribution, we follow the strategy of decomposing integrands into singular and non-singular terms, which lead respectively to singular and non-singular $\co(1)$  contributions to $G_R$.

\subsubsection{Evaluation of $K_{2(i)}$.}

This term involves $L_1(\alpha)$, given in (\ref{L1def}).
We note that $\nu(\al,\beta)$, defined in (\ref{nudef}), is an analytic function of $(\al,\beta)$ for all $(\al,\beta)$ in the relevant range. The integrand of $L_1$ is singular only at the lower limit, with
\[ \omega(\al,\beta)=O((\beta-\thpp+\al)^{-1/2}),\quad \beta\to (\thpp-\al)^+.\]
Thus $L_1(\al)$ is finite for all $\al\in[\frac{\thpp}{2},\thpp]$. Furthermore, $\tau>2\al$ for all $\al\in[\frac{\thpp}{2},\thpp]$, and so $K_{2(i)}$ corresponds to the integral of a continuous function over a finite interval, yielding
\be K_{2(i)} = \co(1).\label{K2i-result}\ee

\subsubsection{Evaluation of $K_{1(ii)}+K_{2(ii)}$.}

We have
\[ K_{2(ii)} = 32\tau H(2\pi-\thpp-\tauep)\tilde{K}_{2(ii)}, \qquad \tilde{K}_{2(ii)}:=\int_{\thpp}^{\tauep/2}\frac{U(\al)}{(\tau^2-4\al^2)^{3/2}}L_2(\al)d\al\]
with $L_2$ as given in (\ref{L2def}), and note that $\al\in[\thpp,{\tauep/2}]$. Then $\omega$ is singular at the lower limit $\beta=\beta_3$ of the integral $L_2$, and is also singular at the upper limit $\beta=\al$ in the limiting case when $\al = \tauep/2$ and $\tauep=2\pi-\theta$. No other singularities arise in the integral. To take account of both singularities, which correspond to $\beta=\al-\theta$ and $\beta=2\pi-\al-\theta$, we define
\be q_1(\al): = \lim_{\beta\to(\al-\thpp)^+}(\beta-\al+\thpp)^{1/2}\omega(\al,\beta) = \left(\frac{\sin\al\sin(\al-\thpp)}{2\sin\thpp}\right)^{1/2}\label{q1def}\ee
and
\be q_2(\al) := \lim_{\beta\to(2\pi-\al-\thpp)^-}(2\pi-\al-\thpp-\beta)^{1/2}\omega(\al,\beta) = \left(\frac{\sin\al|\sin(\al+\thpp)|}{2\sin\thpp}\right)^{1/2}.\label{q2def}\ee
Then we can write
\be \omega(\al,\beta) = \frac{q_1(\al)}{\sqrt{\beta-\al+\thpp}}+\frac{q_2(\al)}{\sqrt{2\pi-\al-\thpp-\beta}}+\omega_2(\al,\beta),\label{om2def}\ee
where, for all $\al$ in the relevant range, $\omega_2(\al,\beta)$ is a continuous function of $\beta$, and $\partial_\al\omega_2(\al,\beta)$ is an integrable function of $\beta$.
By expanding the analytic function $\nu(\al,\beta)$ about the relevant end-point, we can then write
\begin{eqnarray}
L_2(\al)&=& \int_{\al-\thpp}^{\al} \left\{\frac{q_1(\al)}{\sqrt{\beta-\al+\thpp}}[\nu(\al,\al-\thpp)+O(\beta-\al+\thpp)]\right.\nn\\
&&+ \left.\frac{q_2(\al)}{\sqrt{2\pi-\al-\thpp-\beta}}[\nu(\al,2\pi-\al-\thpp)+O(2\pi-\al-\thpp-\beta)]
+\omega_2(\al,\beta)\nu(\al,\beta)\right\} d\beta \nn\\
&=& q_1(\al)\nu(\al,\al-\thpp)\int_{\al-\thpp}^{\al} \frac{d\beta}{\sqrt{\beta-\al+\thpp}}\nn\\
&&+ q_2(\al)\nu(\al,2\pi-\al-\thpp)\int_{\al-\thpp}^{\al} \frac{d\beta}{\sqrt{2\pi-\al-\thpp-\beta}}+ L_{2,2}(\al)\nn\\
&=& 2\sqrt{\thpp}q_1(\al)\nu(\al,\al-\thpp)+2^{3/2}q_2(\al)\nu(\al,2\pi-\al-\thpp)(\sqrt{\pi-\al}-\sqrt{\pi-\thpp/2-\al}))+L_{2,2}(\al), \label{L22def}
\end{eqnarray}
with an obvious definition of a continuously differentiable function $L_{2,2}$. It follows that
\be
\tilde{L}_2(\al):=L_2(\al)+2^{3/2}q_2(\al)\nu(\al,2\pi-\al-\thpp)\sqrt{\pi-\thpp/2-\al}\label{till2def}\ee
is $C^1$ on $[\thpp,\tauep/2]$ for all $\tauep\leq2\pi-\thpp$.
We can then calculate

\begin{eqnarray}
\tilde{K}_{2(ii)}
&=& \int_\thpp^{\tauep/2}\frac{1}{(\tau-2\al)^{3/2}}\frac{U(\al)}{(\tau+2\al)^{3/2}}
\left[\tilde{L}_2(\al)-2^{3/2}q_2(\al)\nu(\al,2\pi-\al-\thpp)\sqrt{\pi-\thpp/2-\al}\right]\nn\\
&=&\int_\thpp^{\tauep/2}\frac{1}{(\tau-2\al)^{3/2}}\left[\frac{U(\tau/2)}{(2\tau)^{3/2}}\tilde{L}_2(\tau/2)+O(\tau-2\al)\right]d\al\nn\\
&&-2^{3/2}\int_\thpp^{\tauep/2}\frac{U(\al)}{(\tau+2\al)^{3/2}}q_2(\al)\nu(\al,2\pi-\al-\thpp)\frac{\sqrt{\pi-\thpp/2-\al}}{(\tau-2\al)^{3/2}}d\al+\co(1)\nn\\
&=&\frac{U(\tau/2)}{(2\tau)^{3/2}}\tilde{L}_2(\tau/2)\int_\thpp^{\tauep/2}\frac{d\al}{(\tau-2\al)^{3/2}}\nn\\
&&-2^{3/2}\frac{U(\tau/2)}{(2\tau)^{3/2}}q_2(\tau/2)\nu(\tau/2,2\pi-\tau/2-\thpp)\int_\thpp^{\tauep/2}\frac{\sqrt{\pi-\thpp/2-\al}}{(\tau-2\al)^{3/2}}d\al
+\co(1)\nn\\
&=&\frac{U(\tau/2)}{(2\tau)^{3/2}}\left\{\tilde{L}_2(\tau/2)\left[\frac{1}{\sqrt{\tau-\tauep}}-\frac{1}{\sqrt{\tau-2\thpp}}\right]\right.\nn\\
&&\left.-2^{3/2}q_2(\tau/2)\nu(\tau/2,2\pi-\tau/2-\thpp)\left[\frac{\sqrt{\pi-\thpp/2-\tauep/2}}{\sqrt{\tau-\tauep}}-\frac1{2^{3/2}}\ln|\tau-2\pi+\thpp|\right]+\co(1)
\right\}\nn\\
&=&\frac{U(\tau/2)}{(2\tau)^{3/2}}\left\{L_2(\tau/2)\sqrt{\tau/2}\ep^{-1}+q_2(\tau/2)\nu(\tau/2,2\pi-\tau/2-\thpp)\ln|\tau-2\pi+\thpp|\right\} + \co(1).
\label{ktil22sol}
\end{eqnarray}

Incorporating the expression for $K_{1(ii)}$ from (\ref{K12def}), we find that
\begin{eqnarray}
\lim_{\ep\to0^+}(K_{1(ii)}+K_{2(ii)})
&=& 16\frac{U(\tau/2)}{\sqrt{2\tau}}q_2(\tau/2)\nu(\tau/2,2\pi-\tau/2-\thpp)\ln|\tau-2\pi+\thpp|H(2\pi-\thpp-\tau) + \co(1)\nn\\
&=&16\frac{U(\pi-\thpp/2)}{\sqrt{2(2\pi-\thpp)}}q_2(\pi-\thpp/2)\nu_0(\pi-\thpp/2)\ln|\tau-2\pi+\thpp|H(2\pi-\thpp-\tau) + \co(1).\label{K12-K22-limit}
\end{eqnarray}
We note the use here of $f(\tau)\ln|\tau-2\pi+\thpp|=f(2\pi-\thpp)\ln|\tau-2\pi+\thpp|+\co(1)$ for a $C^1$ function $f$, and of
\[ \nu(\al,\al) = \nu_0(\al).\]
See (\ref{vdef}), (\ref{v0def}) and (\ref{nudef}) for this last equation.

\subsubsection{Evaluation of $K_2(iii)$.}

The calculation of this term is very similar to that of $K_{2(ii)}$.
We have
\[ K_{2(iii)}=32\tau H(\tauep-2\pi+\thpp)\tilde{K}_{2(iii)},\qquad \tilde{K}_{2(iii)} := \int_{\thpp}^{\pi-\thpp/2}\frac{U(\al)}{(\tau^2-4\al^2)^{3/2}}L_2(\al)d\al\]
with $L_2$ given by (\ref{L2def}). We encounter singularities of $\omega$ at the lower endpoint $\beta=\al-\thpp$ and at the upper end point $\beta=\al$ in the limiting case $\al =\pi-\thpp/2$. Thus we carry out the double end-point expansion of (\ref{om2def}) and write
\begin{eqnarray*}
L_2(\al) &=& q_1(\al)\nu(\al,\al-\thpp)\int_{\al-\thpp}^\al\frac{d\beta}{\sqrt{\beta-\al+\thpp}}
+ q_2(\al)\nu(\al,2\pi-\al-\thpp)\int_{\al-\thpp}^\al\frac{d\beta}{\sqrt{2\pi-\al-\thpp-\beta}} + L_{2,2}(\al)\\
&=&2\sqrt{\thpp}q_1(\al)\nu(\al,\al-\thpp)+2^{3/2}q_2(\al)\nu(\al,2\pi-\al-\thpp)(\sqrt{\pi-\al}-\sqrt{\pi-\thpp/2-\al}) + L_{2,2}(\al),
\end{eqnarray*}
where $L_{2,2}$ is $C^1$ on the domain of $L_2$. It follows that \[\tilde{L}_2(\al)=L_2(\al)+2^{3/2}q_2(\al)\nu(\al,2\pi-\al-\thpp)\sqrt{\pi-\thpp/2-\al}\]
is also $C^1$ on this domain. Thus we can evaluate $K_{2(iii)}$ by expanding $U\tilde{L}_2/(\tau+2\al)^{3/2}$ around the upper end-point of the integral - we note that this guarantees that we capture the most singular and hence dominant contribution to the integral. (Note also the slight difference at this point in comparison to the calculation of $K_{2(ii)}$.)
\[ \frac{U(\al)L_2(\al)}{(\tau+2\al)^{3/2}} = \frac{U(\pi-\thpp/2)}{(\tau+2\pi-\thpp)^{3/2}}\left\{ L_2(\pi-\thpp/2) -2^{3/2}q_2(\pi-\thpp/2)\nu(\pi-\thpp/2,\pi-\thpp/2)\sqrt{\pi-\thpp/2-\al} + O(\pi-\thpp/2-\al)\right\}.\]
Hence
\begin{eqnarray}
\tilde{K}_{2(iii)} &=& \frac{U(\pi-\thpp/2)}{(\tau+2\pi-\thpp)^{3/2}}L_2(\pi-\thpp/2)\int_{\thpp}^{\pi-\thpp/2} \frac{d\al}{(\tau-2\al)^{3/2}}\nn\\
&&-\frac{U(\pi-\thpp/2)}{(\tau+2\pi-\thpp)^{3/2}}2^{3/2}q_2(\pi-\thpp/2)\nu(\pi-\thpp/2,\pi-\thpp/2)\int_{\thpp}^{\pi-\thpp/2} \frac{\sqrt{\pi-\thpp/2-\al}}{(\tau-2\al)^{3/2}}d\al\nn\\
&&+ \int_{\thpp}^{\pi-\thpp/2} \frac{O(\pi-\thpp/2-\al)}{(\tau-2\al)^{3/2}}d\al\nn\\
&=&\frac{U(\pi-\thpp/2)}{(\tau+2\pi-\thpp)^{3/2}}L_2(\pi-\thpp/2)\left\{\frac{1}{\sqrt{\tau-2\pi+\thpp}}-\frac{1}{\sqrt{\tau-2\thpp}}\right\}\nn\\
&&+\frac{U(\pi-\thpp/2)}{(\tau+2\pi-\thpp)^{3/2}}q_2(\pi-\thpp/2)\nu_0(\pi-\thpp/2)\ln|\tau-2\pi+\thpp|\nn\\
&&+O((\tau-2\pi+\thpp)^{1/2})+O\left(\frac{1}{\sqrt{\tau-2\thpp}}\right).
\label{K23-orders}
\end{eqnarray}
Note the use of $\nu(\al,\al)=\nu_0(\al)$. Given that $\tau>\pi>2\thpp$, we can therefore write
\begin{eqnarray} \lim_{\ep\to0^+}K_{2(iii)} &=& \frac{16U(\pi-\thpp/2)L_2(\pi-\thpp/2)}{\sqrt{2(2\pi-\thpp)}\sqrt{\tau-2\pi+\thpp}}H(\tau-2\pi+\thpp)\nn\\
&& +\frac{16U(\pi-\thpp/2)q_2(\pi-\thpp/2)}{\sqrt{2(2\pi-\thpp)}}\nu_0(\pi-\thpp/2)\ln|\tau-2\pi+\thpp|H(\tau-2\pi+\thpp)+\co(1).
\label{K23-limit}
\end{eqnarray}

\subsubsection{Evaluation of $K_{1(iv)}+K_{2(iv)}$.}
We have
\[ K_{2(iv)} = 32\tau H(\tauep-2\pi+\thpp)\tilde{K}_{2(iv)},\qquad \tilde{K}_{2(iv)}:=
\int_{\pi-\thpp/2}^{\tauep/2}\frac{U(\al)}{(\tau^2-4\al^2)^{3/2}}L_3(\al)d\al,\]
with $L_3$ as defined in (\ref{L3def}). The integrand of $L_3$ is singular at the endpoints of the interval of integration, which, in terms of the definitions above, are $\beta_3=\al-\thpp$ and $\beta_1+2\pi=2\pi-\al-\thpp$. We note that $\beta_2<\beta_3$ for the range of values of $\al$ of concern here. Also, with $\al>\pi-\thpp/2$, we have $\al+\thpp>\pi+\thpp/2>\pi$, and so $2\pi-\al-\thpp<\al+\thpp=\beta_4$. Therefore $\beta_3$ and $\beta_1+2\pi$ are the \textit{only} points at which the integrand of $L_3$ is singular. So again we use the double end-point expansion (\ref{om2def}) and write

\begin{eqnarray}
L_3(\al)&=& \int_{\al-\thpp}^{2\pi-\al-\thpp} \left\{\frac{q_1(\al)}{\sqrt{\beta-\al+\thpp}}[\nu(\al,\al-\thpp)+O(\beta-\al+\thpp)]\right.\nn\\
&&+ \left.\frac{q_2(\al)}{\sqrt{2\pi-\al-\thpp-\beta}}[\nu(\al,2\pi-\al-\thpp)+O(2\pi-\al-\thpp-\beta)]
+\omega_2(\al,\beta)\nu(\al,\beta)\right\} d\beta \nn\\
&=& q_1(\al)\nu(\al,\al-\thpp)\int_{\al-\thpp}^{2\pi-\al-\thpp} \frac{d\beta}{\sqrt{\beta-\al+\thpp}}\nn\\
&&+ q_2(\al)\nu(\al,2\pi-\al-\thpp)\int_{\al-\thpp}^{2\pi-\al-\thpp} \frac{d\beta}{\sqrt{2\pi-\al-\thpp-\beta}}\nn\\
&&+ L_{3,2}(\al)\nn\\
&=& 2^{3/2}q_1(\al)\nu(\al,\al-\thpp)(\pi-\al)^{1/2}+2^{3/2}q_2(\al)\nu(\al,2\pi-\al-\thpp)(\pi-\al)^{1/2}+L_{3,2}(\al), \label{L32def}
\end{eqnarray}
with an obvious definition of $L_{3,2}$ By analyticity of $\nu$ and the properties of $\omega_2$ noted above, it follows that $L_{3,2}$ is a continuously differentiable function of $\al$ on $[\pi-\thpp/2,\tau/2]$.

Then we calculate $\tilde{K}_{2(iv)}$ by expanding $L_3(\al)$ about $\al=\tau/2$:
\[ L_3(\al)=L_3(\tau/2)+O(\al-\tau/2),\]
so that
\begin{eqnarray}
\tilde{K}_{2(iv)} &=& \int_{\pi-\thpp/2}^{\tauep/2} \left\{ \frac{U(\tau/2)L_3(\tau/2)}{(2\tau)^{3/2}(\tau-2\al)^{3/2}} + O((\tau-2\al)^{-1/2})\right\} d\al
\nn\\
&=& \frac{U(\tau/2)L_3(\tau/2)}{(2\tau)^{3/2}}\left[\frac{1}{\sqrt{\tau-\tauep}}-\frac{1}{\sqrt{\tau-2\pi+\thpp}}\right] + \co(1).
\end{eqnarray}

Feeding through to $K_{2(iv)}$, we see that the first term here cancels (to $\co(1)$) with $K_{1(iv)}$ in the limit $\ep\to 0^+$, and we find

\begin{eqnarray} \lim_{\ep\to0^+} (K_{1(iv)}+K_{2(iv)}) &=& -\frac{16U(\tau/2)L_3(\tau/2)}{\sqrt{2\tau}\sqrt{\tau-2\pi+\thpp}}H(\tau-2\pi+\thpp) + \co(1)\nn\\
&=& -\frac{16U(\pi-\thpp/2)L_3(\pi-\thpp/2)}{\sqrt{2(2\pi-\thpp)}\sqrt{\tau-2\pi+\thpp}}H(\tau-2\pi+\thpp) + \co(1)
\label{K14+K24-limit}
\end{eqnarray}
We use here $f(\tau)/\sqrt{\tau-2\pi+\thpp} = f(2\pi-\thpp)/\sqrt{\tau-2\pi+\thpp}+\co(1)$ for a function $C^1$ at $2\pi-\thpp$.

\subsubsection{Summary: the $U\times V$ contribution.}

Combining the results (\ref{K2i-result}), (\ref{K12-K22-limit}), (\ref{K23-limit}) and (\ref{K14+K24-limit}) and using
\[ L_2(\pi-\thpp/2) = L_3(\pi-\thpp/2)\]
which can be seen to follow immediately from the definitions (\ref{L2def}) and (\ref{L3def}), we obtain
\be G_2+G_3+G_4 = \frac{16U(\pi-\thpp/2)q_2(\pi-\thpp/2)}{\sqrt{2(2\pi-\thpp)}}\nu_0(\pi-\thpp/2)\ln|\tau-2\pi+\thpp|+\co(1).
\label{G234-result} \ee

\subsection{$G_R(x,\xpp)$ beyond the normal neighbourhood.}

Combining (\ref{g1-6}), (\ref{G1-result}), (\ref{G6-result}), (\ref{G5-result}) and (\ref{G234-result}) yields the expression for $G_R(x,\xpp)$ in the case where $\pi<\tau<2\pi$. This corresponds to the situation where $x$ is outside the maximal normal neighbourhood of $\xpp$, and where the $\st$-envelope of null geodesics emerging from $\xpp$ have passed through a single caustic. We have
\begin{eqnarray}
G_R(x,\xpp)&=&-\frac{1}{\pi\sqrt{\sin\theta}\sqrt{2\pi-\theta}}PV\left(\frac{1}{\tau-2\pi+\theta}\right)\nn\\
&& -\frac{\ln|\tau-2\pi+\theta|}{4\pi\sqrt{2\pi-\theta}}\left\{\frac{16}{\sqrt{2}}U(\pi-\theta/2)q_2(\pi-\theta/2)\nu_0(\pi-\theta/2)-\frac{(2\pi-\theta)(2+\cos\theta)+3\sin\theta}{2(2\pi\theta)(\sin(\theta)^{3/2}}\right\}
+\co(1).
\end{eqnarray}
Using the definitions (\ref{udef}), (\ref{v0def}) and (\ref{q2def}) of $U$, $\nu_0$ and $q_2$ respectively, this yields
\begin{eqnarray}
G_R(x,\xpp)&=&-\frac{1}{\pi\sqrt{\sin\theta}\sqrt{2\pi-\theta}}PV\left(\frac{1}{\tau-2\pi+\theta}\right)\nn\\
&& -\frac{1}{8\pi}\left(\frac{2\pi-\theta}{\sin\theta}\right)^{1/2}\left[1-m^2-4\xi R+\frac{1}{(2\pi-\theta)^2}-\frac{\cot(2\pi-\theta)}{2\pi-\theta}\right]\ln|\tau-2\pi+\theta| +\co(1)\nn\\
&=&-2\left|\frac{2\pi-\theta}{\sin(2\pi-\theta)}\right|^{1/2}PV\left(\frac{1}{\pi(\tau^2-(2\pi-\theta)^2)}\right)-\nu_0(2\pi-\theta)\frac{\ln|\tau^2-(2\pi-\theta)^2|}{\pi}+\co(1).
\end{eqnarray}
We have used here linearity of $PV$, the property that $f(0)PV(1/x)=f(x)PV(1/x)+\co(1)$ and the definition (\ref{v0def}) of $\nu_0$. We have also absorbed an $\co(1)$ term into the logarithm to facilitate the final step. This step is to rewrite the last equation in covariant form. We find
\begin{eqnarray} G_R(x,\xpp)= U(2\pi-\gamma)PV\left(\frac{1}{\pi\sigma(x,\xpp)}\right)-\nu_0(2\pi-\gamma)\frac{\ln|\sigma(x,\xpp)|}{\pi}+\co(1).
\label{GRfinal}
\end{eqnarray}

\subsection{The case $\gamma=\pi$.}
A perplexing feature of the result (\ref{GRfinal}) is that it diverges at $\gamma=\pi$ at points $x$ \textit{not} connected to the base point $\xpp$ by a null geodesic. This can be understood by considering the method by which this result was obtained, and, in particular, the use of the property $f=\co(1)$. Among the various terms swept under the $\co(1)$ carpet are terms involving negative powers of $\sin\theta$ (i.e.\ negative powers of $\sin\gamma$). It is only permissible to ignore these when $\sin\gamma$ is bounded away from zero. Thus the result (\ref{GRfinal}) should be understood with this caveat in mind.

Furthermore, the use of the coordinates $(\al,\beta)$ on $\st$ is not valid when $\gamma=\pi$. This case corresponds to the use of the geodesic distances of a point on $\st$ from the north and south poles as coordinates on $\st$, which is clearly invalid. However we can calculate $G_R(x,\xpp)|_{\gamma=\pi}$ by returning to the calculation above prior to the introduction of these coordinates. To that end, we note that
\[ \gamma_1=\al,\qquad \gamma_2=\beta = \pi-\al,\]
and so
\[ \sigma_1=-\frac18\tau^2+\frac12\yp^2+\frac12\al^2,\quad \sigma_2=-\frac18\tau^2+\frac12\yp^2+\frac12(\pi-\al)^2.\]
As coordinates on $\st$, we use the usual coordinates $\theta^\prime=\al$ and $\phi^\prime$. The calculation of $G_R(x,\xpp)$ is made considerably easier by virtue of the fact that all integrands encountered are independent of $\phi^\prime$. The calculation below assumes (as above) that $\pi<\tau<2\pi$.

To determine $G_1$ in the case $\gamma=\pi$, we take up the calculation leading to (\ref{g12}) at the second line:
\begin{eqnarray} G_1&=&\tau\int_\st\left(\frac{\al}{\sin\al}\right)^{1/2}\left(\frac{\pi-\al}{\sin(\pi-\al)}\right)^{1/2}\left[\left.\frac{\delta^\prime(\sigma_2)}{\yp}\right|_{\yp=y_1}H(\tau-2\al)
+\left.\frac{\delta^\prime(\sigma_1)}{\yp}\right|_{\yp=y_2}H(\tau-2(\pi-\al))\right]\sin\al\,d\al\,d\phi^\prime\nn\\
&=&2\pi\tau\int_0^\pi\sqrt{\al(\pi-\al)}\left[\left.\frac{\delta^\prime(\sigma_2)}{\yp}\right|_{\yp=y_1}H(\tau-2\al)
+\left.\frac{\delta^\prime(\sigma_1)}{\yp}\right|_{\yp=y_2}H(\tau-2(\pi-\al))\right]\,d\al \label{G1pi1}
\end{eqnarray}

We have
\[ \delta^\prime(\sigma_1)|_{\yp=y_2}=\frac{1}{\pi^2}\frac{d}{d\al}\left\{\delta(\al-\pi/2)\right\},\qquad \delta^\prime(\sigma_2)|_{\yp=y_1}=-\frac{1}{\pi^2}\frac{d}{d\al}\left\{\delta(\al-\pi/2)\right\}.\]
Using these and integrating by parts gives
\begin{eqnarray}
G_1 &=&\frac{4\tau}{\pi}\int_0^\pi \frac{d}{d\al}
\left\{
\frac{\sqrt{\al(\pi-\al)}}{\sqrt{\tau^2-4\al^2}}H(\tau-2\al)-\frac{\sqrt{\al(\pi-\al)}}{\sqrt{\tau^2-4(\pi-\al)^2}}H(\tau-2(\pi-\al))\right\}\delta(\al-\pi/2)d\al\nn\\
&=&\frac{4\tau}{\pi}\left.\frac{d}{d\al}
\left\{
\frac{\sqrt{\al(\pi-\al)}}{\sqrt{\tau^2-4\al^2}}H(\tau-2\al)-\frac{\sqrt{\al(\pi-\al)}}{\sqrt{\tau^2-4(\pi-\al)^2}}H(\tau-2(\pi-\al))\right\}\right|_{\al=\pi/2}\nn\\
&=&-\frac{8\tau}{(\tau^2-\pi^2)^{1/2}}\delta(\tau-\pi)+\frac{8\pi\tau}{(\tau^2-\pi^2)^{3/2}}H(\tau-\pi).\label{G1pires}
\end{eqnarray}

Next, we recall that
\[ G_2+G_3+G_4 = \lim_{\ep\to0^+}G_{2,3,4}^{(\ep)},\]
as given by (\ref{g34epk}). From (\ref{K1def}), with $\gamma=\pi$ (giving $\beta=\al-\pi$ and $H(\al-\beta)\equiv1$), we have
\begin{eqnarray}
K_1 &=&-2\tau\int_\st\left(\frac{\al}{\sin\al}\right)^{1/2}\left.\frac{V_2}{\ep}\right|_{\yp=\ep}\delta\left(-\frac{\tau^2}{8}+\frac{\ep^2}{2}+\frac{\al^2}{2}\right)\sin\al\,d\al\,d\phi^\prime\nn\\
&=& -\frac{8\pi\tau}{\tauep}\int_0^\pi (\al\sin\al)^{1/2}\left.\frac{V_2}{\ep}\right|_{\yp=\ep}\delta(\al-\tauep/2)d\al\nn\\
&=&-\frac{8\pi\tau}{\ep\tauep}\left(\frac{\tauep}{2}\sin(\tauep/2)\right)^{1/2}\sum_{k=0}^\infty\nu_k(\pi-\tauep/2)\left(\frac{\pi}{2}(\pi-\tauep)\right)^k.\label{K1pires}
\end{eqnarray}

Likewise, from (\ref{K2def}), we obtain
\begin{eqnarray}
K_2&=&32\pi\tau\int_0^\pi\frac{(\al\sin\al)^{1/2}}{(\tau^2-4\al^2)^{3/2}}V_2|_{\yp=y_1}H(\tauep/2-\al)\,d\al\nn\\
&=&32\pi\tau\int_0^{\tauep/2}\frac{(\al\sin\al)^{1/2}}{(\tau^2-4\al^2)^{3/2}} \sum_{k=0}^\infty \nu_k(\pi-\al)(\frac{\pi}{2}(\pi-2\al))^k\,d\al
\end{eqnarray}

We expand
\[ Q(\al)=(\al\sin\al)^{1/2}\sum_{k=0}^\infty \nu_k(\pi-\al)\left(\frac{\pi}{2}(\pi-2\al)\right)^k\]
around $\al=\tau/2$ and integrate. The leading order term is proportional to $\ep^{-1}$, and cancels identically with $K_1$. The remaining terms are finite, and so
\[ K_1+K_2 = \co(1),\]
leading to a finite value for $G_2+G_3+G_4$ in the case when $\gamma=\pi$.

The analysis of $G_5+G_6$ done in subsection C carries over to the case $\gamma=\pi$, and so
\[ G_5+G_6=\co(1)\]
also holds in this case.

Combining these results, we find
\begin{eqnarray} G_R(x,\xpp)|_{\gamma=\pi} &=& \frac{2\tau}{\pi}\left\{\frac{1}{(\tau^2-\pi^2)^{1/2}}\delta(\tau-\pi)-\frac{\pi}{(\tau^2-\pi^2)^{3/2}}H(\tau-\pi)\right\} + \co(1),\qquad (\tau<2\pi)\nn\\
&=&\frac{2\tau}{\pi}\left\{\frac{d}{d\tau}\left[\frac{H(\tau-\pi)}{(\tau^2-\pi^2)^{1/2}}\right]+\frac{H(\tau-\pi)}{(\tau-\pi)^{1/2}(\tau+\pi)^{3/2}}\right\} + \co(1),\qquad (\tau<2\pi).
\label{GRpires}\end{eqnarray}

Note that the final form here shows explicitly the distributional nature of the expression, which is not manifest in the previous line. 
Of course for $\tau<\pi$, the $\co(1)$ term is identically zero.

This result shows that the only singularity arising at $\gamma=\pi$ is confined to the null cone (as expected). This highlights a drawback of the result (\ref{GRfinal}): it does not apply on a neighbourhood of $\gamma(x,\xpp)=\pi$.


\section{Singularity structure via mode sum approximation}\label{sec:Large l}

In this section, we present the results of a completely different approach to the calculation of the singular part of the retarded Green's function outside the normal neighbourhood. This approach - involving asymptotic approximations for the special functions that arise in a mode sum decomposition of $G_R$ - has the advantage of yielding global results. That is, we obtain the singular part of $G_R$ globally. The drawbacks are that we must restrict to a particular value of the coupling term $m^2+\xi R$, and we obtain results which have unresolved causality issues. (It is clear however that these issues must be resolvable, and they do not affect the correct identification of the singular part of $G_R$.) The results are of interest in that they are global in nature: they are not restricted either to the normal neighbourhood or to the region before the formation of the second caustic. What is of most interest is that they return the four-fold singularity structure discussed above.

To begin, we find the Green's function solutions of the wave equation (\ref{weGret}) by carrying out
a multipole decomposition of a solution of the wave equation (\ref{wave}). Imposing the appropriate boundary conditions
we find that the Feynman~\cite{DeWitt:1960,Birrell:Davies} and retarded Green's functions can respectively be expressed as
\begin{equation} \label{eq:G_F mode sum}
G_{F}(x,x')=\lim_{\ep\to 0^+}\frac{1}{2}\sum_{\ell=0}^{\infty}(\ell+1/2)H_0^{(2)}\left(\lambda \etae\right)P_{\ell}(\cos\gamma),
\end{equation}
and
\begin{equation} \label{eq:Gret on M_2xS^2}
G_R(x,x')=2\textrm{Re}(G_F(x,x'))H_+(x,\xp)=H_+(x,\xp)\sum_{\ell=0}^{\infty}(\ell+1/2)J_0\left(\lambda\etaM\right)P_{\ell}(\cos\gamma).
\end{equation}

The expression (\ref{eq:G_F mode sum}) is obtained by expanding $G_F$ in terms of Legendre polynomials, and applying equation (2.77) of \cite{Birrell:Davies} to obtain the resulting (1+1)-dimensional Feynman Green's function. In these expressions, $\etae:= \sqrt{\left(|t-\tp|-i\epsilon\right)^2-(y-\yp)^2}$ and $\lambda^2:= \ell(\ell+1)+m^2+2\xi$. 

We find it more useful to carry out the large-$\ell$ asymptotic analysis on the Feynman Green function and then
obtain from it the large-$\ell$ analysis for the retarded Green function.
Therefore, we expand Eq.(\ref{eq:G_F mode sum}) for large-$\ell$ - i.e.\ large argument for the Hankel function $H_0^{(2)}$ - order-by-order.
For that we make use of Eq.8.451(4) of~\cite{GradRyz}:
\begin{equation}
H_{0}^{(2)}(z)=\left(\frac{2}{\pi z}\right)^{1/2}e^{-i(z-\pi/4)}
\left[
\sum_{k=0}^{n-1}
\frac{\alpha_k}{z^k}+\theta_2\frac{\alpha_n}{z^n}
\right],\qquad
\alpha_k\equiv \frac{1}{(2i)^k}\frac{\Gamma(k+1/2)}{k!\Gamma(-k+1/2)}
\end{equation}
where (\cite{GradRyz}, p.\ 963) $\left|\theta_2\right|<1$ if $-3\pi/2<\arg(z)<\pi/2$ and $\text{Im}(z)\leq 0$, which is the case here:
\[ z = \left(\ell+\frac12\right)\etae \sim \left(\ell+\frac12\right)\eta - i\left(\ell+\frac12\right)\frac{\Delta t}{\eta}\epsilon,\qquad \epsilon\to0^+.\]
$\text{Im}(z)\leq 0$ then arises as we are only concerned with the case $\eta>0$ (and note that $\Delta t=|t-\tp|$). We have a contribution to $G_R$ only when $\sigma\leq 0$, so that $\eta\geq 0$. But $\eta=0$ with $\sigma\leq 0$ can only arise when $x=\xp$, which case can be ignored.

We take $n=2$, giving
\be H_{0}^{(2)}(z)=\left(\frac{2}{\pi z}\right)^{1/2}e^{-i(z-\pi/4)}\left[1+\frac{i}{8z}-\frac{9}{2^7}\theta_2z^{-2}\right].\label{hankn2}\ee

Consider the remainder term $G_F^{\textrm{rem}}$ - i.e.\ the contribution of $\theta_2$ to $G_F$. This yields
\be G_F^{\textrm{rem}} =
 \lim_{\ep\to 0^+}\frac{9}{2^8}\sqrt{\frac{2}{\pi}}\etae^{-5/2}e^{-i\etae/2}\sum_{\ell=0}^\infty \theta_2\left(\ell+\frac12\right)^{-3/2}e^{-i\ell\etae}P_\ell(\cos\gamma).\ee
Since $|\theta_2|<1$, we have
\[ \left|\sum_{\ell=0}^\infty \theta_2(\ell+\frac12)^{-3/2}e^{-i\ell\etae}P_\ell(\cos\gamma)\right|\leq \sum_{\ell=0}^\infty\lh^{-3/2},\]
so that the series is absolutely convergent. Furthermore, each term in the series is continuous on $\Omega=\{(z,\gamma):|z|\leq 1, \gamma\in[0,\pi]\}$, and so the Weierstrass $M-$test yields uniform convergence to a continuous function. Therefore $G_F^{\textrm{rem}}$ is continuous for $\etae\neq0$. It follows that the singularities of $G_F$ and $G_R$ are contained in the terms
\begin{eqnarray}
G_F^{(0)} &:=& -\lim_{\ep\to 0^+}\frac12\sqrt{\frac{2}{\pi}}e^{i\pi/4}\etae^{-1/2}e^{-i\etae/2}S_0,\label{gf0def}\\
G_F^{(1)} &:=& -\lim_{\ep\to 0^+}\frac{i}{16}\sqrt{\frac{2}{\pi}}e^{i\pi/4}\etae^{-3/2}e^{-i\etae/2}S_1,\label{gf1def}
\end{eqnarray}
where
\be S_k:=\lsumo\lh^{1/2-k}e^{-i\ell\etae}\plg,\qquad k=0,1\label{Skdef}\ee
and the (exact) decomposition
\[ G_F = G_F^{(0)} + G_F^{(1)} + G_F^{\textrm{rem}}\]
holds.

To proceed, we note that the summands of both $S_k, k=0,1$ are continuous and bounded functions on their relevant domains. Hence any singularity that arises does so because of the divergence (as functions) of the infinite series. In other words, the divergence is due to the `large-$\ell$' contributions to $S_k$. To account for these, we can use large-$\ell$ approximations for the Legendre polynomials $\plg$. We note first the trivial results that
\[ S_0 = \frac{1}{\sqrt{2}}+T_0,\qquad S_1=\sqrt{2}+T_1,\]
where
\be T_k:=\lsumi\lh^{1/2-k}e^{-i\ell\etae}\plg,\qquad k=0,1.\label{Tkdef}\ee
Then we apply the Bonnet-Heine formula (see~\cite{Sansone}, p.\ 208):
\be \plg = \plgt +O(\ell^{-5/2}).
\label{bonnet-heine1}
\ee
where
\be \plgt = \sqrt{\frac{2}{\pi\ell\sin\gamma}}\left\{\left(1-\frac{1}{4\ell}\right)\cos[\lh\gamma-\pi/4]+\frac{1}{8\ell}\cot\gamma\sin[\lh\gamma-\pi/4]\right\}.
\label{bonnet-heine2}
\ee
This relation holds for $\epsilon\leq\gamma\leq\pi-\epsilon$, $0<\epsilon<\pi/2$. We note that, as above, the remainder contributes an overall continuous term to $T_k, k=0,1$. It follows that the singular contributions to $G_F$ arise solely from
\be
\tilt_k:=\lsumi\lh^{1/2-k}e^{-i\ell\etae}\plgt,\qquad k=0,1.\label{tiltkdef}
\ee

%
%

For convenience, we introduce the notation
\[ A\doteq B\]
to mean that $A-B$ is a continuous function. This gives a convenient way of indicating the removal of different continuous contributions to the various quantities encountered. We note also that in each case, continuous functions arising from infinite sums are identified by repeating the $M-$test/uniform convergence argument used above. In practice, this simply amounts to removing terms of order $\ell^{-2}$ from the relevant series. That is, to obtain the singular parts of $\tilt_k$, we expand the non-oscillatory factors in inverse powers of $\ell$, and discard all $O(\ell^{-2})$ terms. These do not contribute to the discontinuous part of the retarded Green's function.

Thus gathering relevant terms, we have

\begin{eqnarray}
\tilt_0 &\doteq& -\frac{i}{8\sqrt{2\pi}}\frac{\cot\gamma}{\sqrt{\sin\gamma}}\left\{e^{i(\gamma/2-\pi/4)}\ca(\gamma-\etae)-e^{-i(\gamma/2-\pi/4)}\ca(-(\gamma+\etae))\right\}\nn\\
&& + \frac12\sqrt{\frac{2}{\pi\sin\gamma}}\left\{e^{i(\gamma/2-\pi/4)}\cb(\gamma-\etae)+e^{-i(\gamma/2-\pi/4)}\cb(-(\gamma+\etae))\right\},\label{tilt0result}\\
\tilt_1 &\doteq& \frac12\sqrt{\frac{2}{\pi\sin\gamma}}\left\{e^{i(\gamma/2-\pi/4)}\ca(\gamma-\etae)+e^{-i(\gamma/2-\pi/4)}\ca(-(\gamma+\etae))\right\},\label{tilt1result}
\end{eqnarray}

where

\be \ca(z) := \lsumi\ell^{-1}e^{i\ell z},\qquad \cb(z) := \lsumi e^{i\ell z}.\label{calabdef}\ee

Recall that in order to obtain the retarded Green's function, we require the limit $\epsilon\to 0^+$ of the terms above. This means that we need only evaluate $\ca$ and $\cb$ for real arguments: these have well-defined distributional forms. We give the results here, leaving the relevant derivations to Appendix B. For $s\in\mathbb{R}$, we find

\be \ca(s)=\frac12\cd(s)+i\cu(s),\label{caform}\ee
where
\begin{eqnarray}
\cd(s) &=& -2\ln|s|-2\sum_{k=0}^\infty\ln\left|1-\frac{s^2}{4k^2\pi^2}\right|,\label{cdform}\\
\cu(s) &=& \frac12(\pi-s)+\pi\left\{\sum_{k=1}^\infty H(s-2k\pi)-\sum_{k=0}^\infty H(-s-2k\pi)\right\}\label{cuform}
\end{eqnarray}
and
\be \cb(s) = -\frac{i}{2}\cd^\prime(s)+\cu^\prime(s),\label{cbform}\ee
with
\begin{eqnarray}
\cd^\prime(s) &=& -2\sum_{k\in\mathbb{Z}}PV\left(\frac{1}{s+2k\pi}\right),\label{cdprimeform}\\
\cu^\prime(s) &=& -\frac12+\pi\sum_{k\in\mathbb{Z}}\delta(s-2k\pi).\label{cuprimeform}
\end{eqnarray}
Derivatives here are distributional, as indeed are the expressions themselves. For example, $\cd$ is defined almost everywhere on $\mathbb{R}$ and is locally integrable, and so yields a distribution.

Recall that our aim is to determine $G_R$, which is obtained by taking the real part of $G_F$, the calculation of which involves taking the limit $\epsilon\to 0^+$ of $\ca(z)$ and $\cb(z)$. These functions are analytic on the lower half plane, which is precisely the situation that holds here: the arguments of $\ca, \cb$ that we encounter have the form
\[ z = \pm\gamma-\etae \sim \pm\gamma-\eta + i\frac{\Delta t}{\eta},\qquad \epsilon\to 0^+.\] This means that to determine $G_R$, we can simply set $\epsilon=0$ in the expressions above that combine to give $G_F$, and then take the real part (with the appropriate constant factor). The resulting terms are distributions rather than analytic functions.

Of course we are not calculating the full $G_R$, only its discontinuous part which we denote $\grsing$. This is constructed from terms of the form $\ca(\gamma-\eta)$ and $\cb(\gamma-\eta)$ (respectively $\ca(-(\gamma+\eta))$ and $\cb(-(\gamma+\eta))$, scaled by factors that include terms $e^{i(\gamma-\eta)/2}$ (respectively $e^{i(\gamma-\eta)/2}$). This can be verified by tracking through equations (\ref{gf0def}) to (\ref{tilt1result}). Determining the final result for the singular part of $G_R$ is simplified by the following observation: the singularities of $\ca(\gamma-\eta)$ are concentrated at points with $\gamma-\eta=2k\pi, k\in\mathbb{Z}$. At these points, $\sin((\gamma-\eta)/2)=0$. This factor mollifies the singular behaviour, and renders the corresponding term continuous. The possible combinations are essentially functions of the form $s\delta(s)$, $sPV(\frac{1}{s})$, $s\ln s$ and $s H(s)$, each one of which is continuous. The same holds for the singularities of $\cb$, and when the argument is $-(\gamma+\eta)$. A consequence of this, which simplifies the remainder of the calculation, is that there is no contribution to $\grsing$ from terms with coefficients of the form $\sin((\gamma\pm\eta)/2)$. Likewise,
\[ \cos\left(\frac{\gamma\pm\eta}{2}\right)\sum_{k}\Delta(\gamma\pm\eta-2k\pi) \doteq \sum_k (-1)^k \Delta(\gamma\pm\eta-2k\pi),\]
where $\Delta(s)$ is any one of $\delta(s), PV(\frac{1}{s}), \ln|s|$ or $H(s)$ with singularities at $s=2k\pi$.

With these observations in hand, we can determine $\grsing$ (for convenience, we do not include the factor $H_+(x,\xp)$:
\begin{eqnarray}
\grsing &\doteq& \frac1\pi\eta^{-1/2}\frac{1}{\sqrt{\sin{\gamma}}} \times \nn\\
&&+\left\{\cos\left(\frac{\gamma-\eta}{2}\right)\left[\pi\sum_{k\in\mathbb{Z}}\delta(\gamma-\eta-2k\pi)+\frac18\left(\cot\gamma-\frac1\eta\right)\cu(\gamma-\eta)\right]\right.\nn\\
&&+\left.\cos\left(\frac{\gamma+\eta}{2}\right)\left[\sum_{k\in\mathbb{Z}}PV\left(\frac{1}{\gamma+\eta+2k\pi}\right)+\frac{1}{8}\left(\cot\gamma-\frac{1}{\eta}\right)\left(\ln|\eta+\gamma|+
\sum_{k=1}^\infty\ln\left|1-\frac{(\gamma+\eta)^2}{4k^2\pi^2}\right|\right)\right]\right\}.
\label{grsingresult1}
\end{eqnarray}

The following observations yield the final result. First, we incorporate the comment above regarding the factors $\cos((\gamma\pm\eta)/2)$. Next, we apply a similar observation to the term $\eta^{-1/2}$ so that $\eta\to\gamma+2k\pi$ in the $\delta$ and $\cu$ contributions, and $\eta\to2k\pi-\gamma$ in the $PV$ and $\ln$ contributions. Note that in the series $\sum\delta$, there is no contribution from terms with $k\geq1$, for then the argument is always negative. We then apply $\delta(s)=\delta(-s)$ and retain the appropriate sum. A similar argument leads to a truncation of the series implicit in $\cu$ and the $\sum PV$ term. Factors of the argument of the logarithm can also be removed. The coefficients of the singular terms in the series can then be identified: see (\ref{udef}) and (\ref{v0def}). Carrying out these steps yields

\begin{eqnarray}
\grsing &\doteq& \sum_{k=0}^\infty (-1)^k\left[\frac{U(\gamma+2k\pi)}{\gamma+2k\pi}\delta(\eta-(\gamma+2k\pi))+\nu_0(\gamma+2k\pi)H(\eta-(\gamma+2k\pi))\right]\nn\\
&&+\sum_{k=1}^\infty\frac{(-1)^k}{\pi}\left[\frac{U(2k\pi-\gamma)}{2k\pi-\gamma}PV\left(\frac{1}{\eta-(2k\pi-\gamma)}\right)+\nu_0(2k\pi-\gamma)\ln|\eta-(2k\pi-\gamma)|\right].
\label{grsingfinala}
\end{eqnarray}

Next, we note the following results that derive from elementary properties of the distributions involved, and on sign properties of the arguments:
\begin{eqnarray*} \frac{1}{\gamma+2k\pi}\delta(\eta-(\gamma+2k\pi)) &=& \delta(-\frac12\eta^2+(\gamma+2k\pi)^2),\\
H(\eta-(\gamma+2k\pi))&=&H(\frac12\eta^2-\frac12(\gamma+2k\pi)^2),\\
\frac{1}{2k\pi-\gamma}PV\left(\frac{1}{\eta-(2k\pi-\gamma)}\right)&\doteq& PV\left(\frac{1}{\frac12\eta^2-\frac12(2k\pi-\gamma)^2}\right),\\
\ln|\eta-(2k\pi-\gamma)|&\doteq& \ln|\frac12\eta^2-\frac12(2k\pi-\gamma)^2|.
\end{eqnarray*}

Apply these in (\ref{grsingfinala}) yields our final result:

\begin{eqnarray}
\grsing &\doteq& \sum_{k=0}^\infty (-1)^k\left[U(\gamma+2k\pi)\delta(\sigma_k^{\rm{even}})+\nu_0(\gamma+2k\pi)H(-\sigma_k^{\rm{even}})\right]\nn\\
&&+\sum_{k=1}^\infty\frac{(-1)^k}{\pi}\left[U(2k\pi-\gamma)PV\left(\frac{1}{-\sigma_k^{\rm{odd}}}\right)+\nu_0(2k\pi-\gamma)\ln|\sigma_k^{\rm{odd}}|\right],
\label{grsingfinal}
\end{eqnarray}

where
\begin{eqnarray}
\sigma_k^{\rm{even}} &:=& -\frac12\eta^2 + \frac12 (\gamma+2k\pi)^2 = \sigma|_{\gamma\to\gamma+2k\pi},\\
\sigma_k^{\rm{odd}} &:=& -\frac12\eta^2 + \frac12 (2k\pi-\gamma)^2= \sigma|_{\gamma\to 2k\pi-\gamma}.
\end{eqnarray}

The designations even/odd are used to indicate that $\sigma_k^{\rm{even (odd)}}=0$ along a null geodesic that has passed through an even (odd) number of caustic points.

We emphasize that the result (\ref{grsingfinal}) is exact in the sense that
\[ G_R  = (\grsing + G_R^{\textrm{rem}})H_+(x,\xp),\]
where $G_R^{\textrm{rem}}$ is a continuous function. The result applies for values of $\gamma$ bounded away from $0$ and $\pi$ (see the comment immediately following (\ref{bonnet-heine2}).  As noted earlier, caustics form at spacetime points with $\eta=\gamma+2k\pi, k\geq0$ and $\eta=2k\pi-\gamma, k\geq1$. The result (\ref{grsingfinal}) thus reproduces exactly the four-fold singularity structure discussed in the introduction, yielding both the four-fold sequence for the sharp term
\[ \delta(\sigma) \to PV\left(\frac{1}{\sigma}\right) \to - \delta(\sigma) \to -PV\left(\frac{1}{\sigma}\right) \to \delta(\sigma) \to \cdots\]
and the tail term:
\[ \theta(-\sigma) \to -\ln|\sigma| \to -\theta(-\sigma) \to \ln|\sigma| \to \theta(-\sigma) \to \cdots\]
Furthermore, this result yields the coefficients of the singular terms throughout the spacetime.

\begin{figure}[h!]
\begin{center}
\includegraphics[width=3cm]
{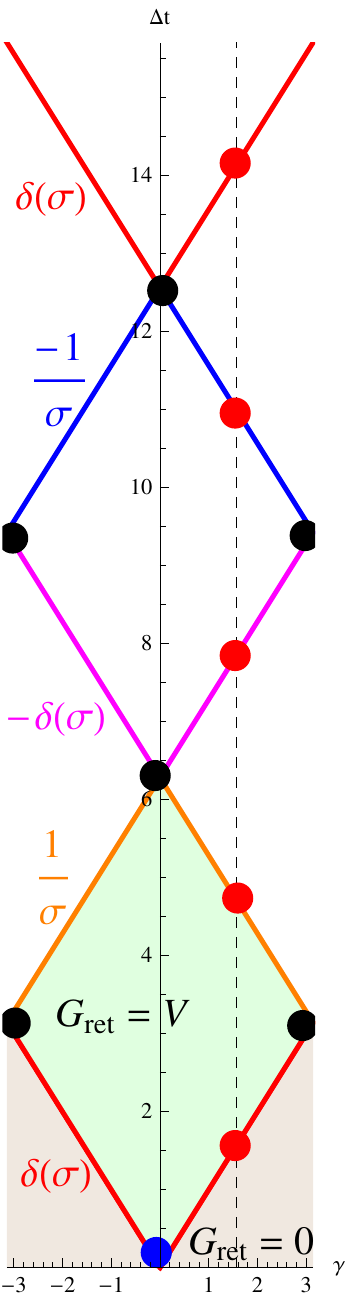}
\includegraphics[width=14cm]
{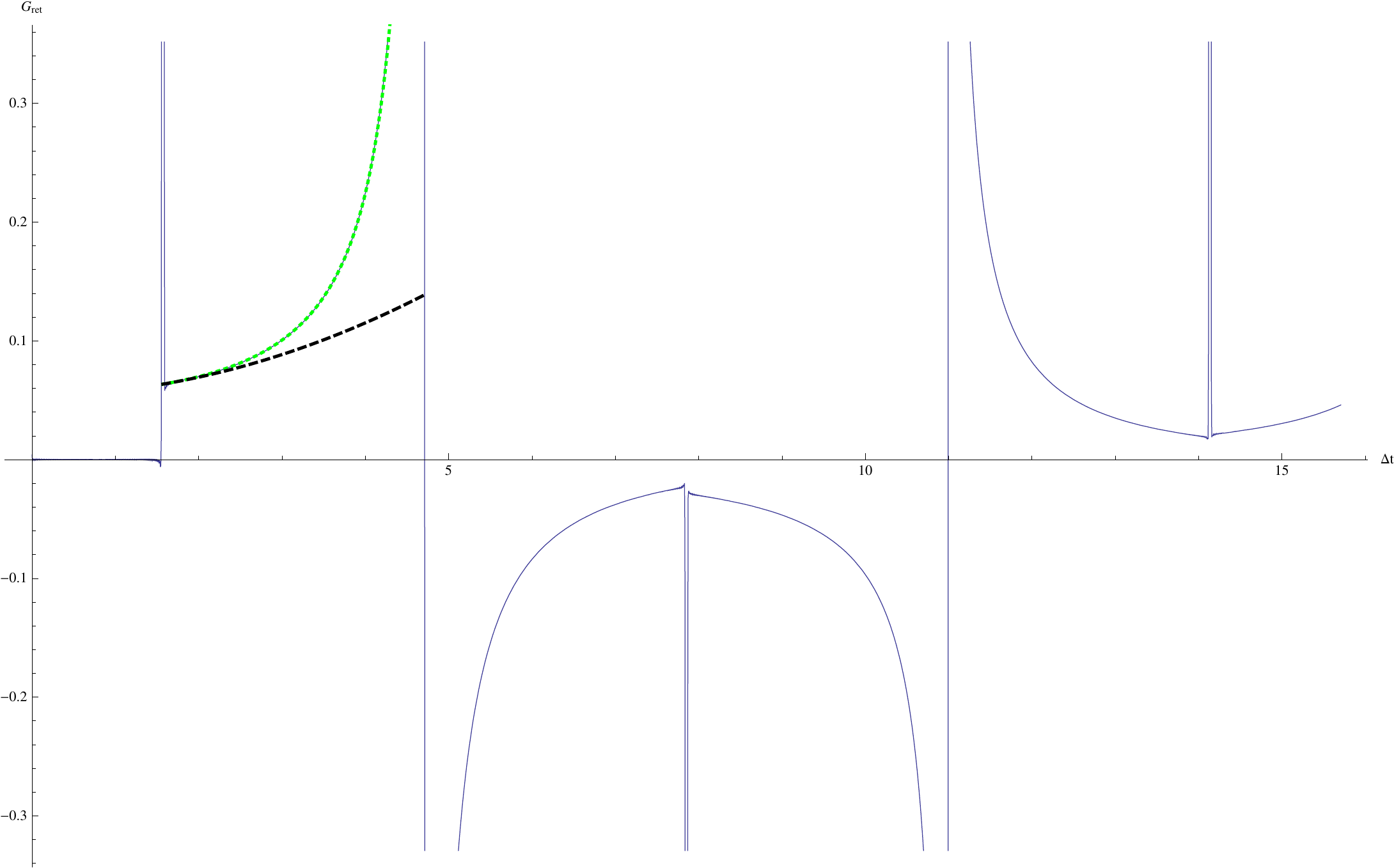}
\end{center}
\caption{
(Colour online.)
(a) (Left panel.) Visualization of PH  spacetime.
Here we have $y=y'$, the time coordinate $T$ runs along the vertical axis and the angular coordinate $\gamma$ runs along the horizontal
axis (which also includes $-\gamma$).
The dashed black vertical line corresponds to $\gamma=\pi/2$: this is a static path.
Black dots on the left and right boundaries correspond to caustic points (for which $\gamma=0,\pi$),
red dots along the dashed line to points where null geodesics intersect the static path at $\gamma=\pi/2$ and the
blue dot on the $\gamma$-axis to the base point $x$.
Diagonal lines correspond to null geodesics; these alternate in colour every time the singularity of $G_{ret}$ changes its character, which is indicated alongside
with a $\pm\delta(\sigma), \pm\frac{1}{\sigma}$.
The brown-shaded region on the bottom corners corresponds to the acausal part of the maximal normal neighbourhood, where $G_{R}=0$.
The green-shaded region on the bottom diamond corresponds to the causal part of the maximal normal neighbourhood, where $G_{R}=V$
(the bottom two diagonal lines, which emanate from $x$, are also contained in  the causal part of the maximal normal neighbourhood, but it is 
instead $G_R=U\delta(\sigma)$ along these lines).
\\
(b) (Right panel.) $G_{R}(x,x')$ as a function of $\dt$ for $y=y'(=0)$,
$\theta=0$, $\phi'=0$, $\thp=\pi/2$.
The blue continuous curve corresponds to the exact $G_R$
obtained using the `smoothed sum method' of Sec.VII.D~\cite{CDOWa}
(such `smoothing' is justified in~\cite{Hardy})
for the mode sum Eq.(\ref{eq:Gret on M_2xS^2})
with $m=0$, $\xi=1/8$, $l_{max}=800$, $l_{cut}=200$.
The dotted green curve is $V(x,x')$ calculated using a WKB Taylor series as in~\cite{CDOWb}.
The dashed black curve is $\nu_0+\nu_1 \sigma$ from Eqs. (\ref{v0def}) and (\ref{eq:hat V1 M2xS2}).
The time for the red dots in (a) correspond to the singularity times of $G_{R}$ in (b). The four-fold
singularity structure
$\delta(\sigma), \text{P.V.}\left(1/\pi\sigma\right), -\delta(\sigma), - \text{P.V.}\left(1/\pi\sigma\right), \delta(\sigma), \dots$ can be clearly seen.
}
\label{fig:Kirchhofff unwrapped cylinder}
\end{figure}

Fig.\ref{fig:Kirchhofff unwrapped cylinder} shows a plot of $G_{R}$ obtained using the $\ell$-mode sum
Eq.(\ref{eq:Gret on M_2xS^2}).
The four-fold singularity structure $\delta(\sigma), \text{P.V.}\left(1/\pi\sigma\right), -\delta(\sigma), - \text{P.V.}\left(1/\pi\sigma\right), \delta(\sigma), \dots$ is clearly manifest in it. The $\ell$-mode sum is also compared with the Hadamard form in the normal neighbourhood.

\section{Discussion}\label{sec:conclusions}

In~\cite{CDOWa}, the following heuristic explanation for the singularity structure of the `direct' term was given involving the Feynman propagator.
The Hadamard form, valid for $\xp\in \normnbd$$(x)$, for the Feynman propagator is~\cite{DeWitt:1960}
\begin{equation}
G_F(x,x')=\lim_{\epsilon \rightarrow 0^+}\frac{i}{2\pi}\left[\frac{U(x,x')}{\sigma+i\epsilon}-V(x,x')\ln\left(\sigma+i\epsilon\right)+W(x,x')\right]
\end{equation}
where the biscalars $U(x,x')=\Delta^{1/2}(x,x')$, $V(x,x')$ are the same as those appearing in the Hadamard form (\ref{grhad}) for the retarded Green function
 and $W(x,x')$ -- like  $U(x,x')$ and $V(x,x')$ -- is a regular and real-valued bitensor defined in a normal neighbourhood of $x$.
Using $G_R(x,x')=2H_+(x,x')\text{Re}\left(G_F(x,x')\right)$
together with
$\lim_{\epsilon \rightarrow 0^+}1/(\sigma + i \epsilon) =\text{P.V.}\left({\frac{1}{\sigma}} \right)-i\pi \delta(\sigma)$
and  $\lim_{\epsilon \rightarrow 0^+} \ln\left(\sigma+i\epsilon\right)=\ln|\sigma|+i\pi\theta(-\sigma)$, it follows that the Hadamard form for the retarded Green's is as given in (\ref{grhad}) above, and is valid for $\xp\in{\cal{N}}(x)$.

In~\cite{CDOWa} it was argued, using the transport equation along the appropriate geodesic joining $x$ and $x'$
which the van Vleck determinant satisfies, that this biscalar
picks up a phase `$-\pi$' every time that the geodesic
passes through a point such that $\gamma=0$ or $\pi$.
In the case of null geodesics, these are caustic points.
The previous argument was given for a static, spherically-symmetric spacetime and this is also seen in~\cite{Casals:Galley} using a completely different method.
Applying then $G_R(x,x')=2H_+(x,x')\text{Re}\left(G_F(x,x')\right)$ to
the `direct' term $\Delta^{1/2}/(\sigma+i\epsilon)$ of the Hadamard form,
the four-fold singularity structure
$\left|\Delta\right|^{1/2}\delta(\sigma), \left|\Delta\right|^{1/2}\text{P.V.}\left(1/\pi\sigma\right), -\left|\Delta\right|^{1/2}\delta(\sigma), - \left|\Delta\right|^{1/2}\text{P.V.}\left(1/\pi\sigma\right), \left|\Delta\right|^{1/2}\delta(\sigma), \dots$
for $G_R$ then follows.

One could apply a similar heuristic argument to the `tail' term $V\ln(\sigma+i\epsilon)$:
considering that $V(x,x')=\Delta^{1/2}(x,x')\bar{V}(x,x')$ with $\bar{V}$ always real-valued (therefore $V$ would pick up the same phase as $U$ after a crossing
of a point where $\gamma=0$ or $\pi$)
it would follow that the `tail' term in the retarded Green's function
should alternate as: $\left|\Delta\right|^{1/2}\bar{V}\theta(-\sigma)$, $-\left|\Delta\right|^{1/2}\bar{V}\ln\left|\sigma\right|/\pi$,
$-\left|\Delta\right|^{1/2}\bar{V}\theta(-\sigma)$, $\left|\Delta\right|^{1/2}\bar{V}\ln\left|\sigma\right|/\pi$, etc.

Using the Kirchhoff integral method we have proven that the suggested alternation after the first caustic-crossing,
both for  the `direct' and the `tail' parts,  is indeed what occurs.
The large-$\ell$ asymptotics of Sec.\ref{sec:Large l} give a completely independent verification of this result and extend it to an arbitrary number of caustic-crossings: the calculations required to apply the Kirchhoff method to evaluate $G_R$ beyond the second caustic become considerably more involved. The large-$\ell$ asymptotics, however, raise a question regarding the causal character of the retarded Green function, which the Kirchhoff integral method
successfully resolves.
The extension of the meaning of the world function $\sigma$, as well as that of the biscalars $U$ and $V$, outside the maximal normal
neighbourhood is cleanly given by Eqs.(\ref{GRresult}), (\ref{sigma1forGR}), and (\ref{grsingfinal}).

Both four-fold cycles -- for the `direct' and the `tail' parts -- together provide a full-account of the singularity structure of the retarded Green's function beyond the normal neighbourhood. They are proven here for the PH spacetime
but we speculate that this is true for a general set of spacetimes which includes Schwarzschild.
Both methods,  the Kirchhoff integral method and the large-$\ell$ asymptotics method, are in principle applicable to Schwarzschild spacetime.
Although the Kirchhoff integral method seems to be of difficult application to Schwarzschild, we hope to be able to apply
the large-$\ell$ asymptotics method to Schwarzschild.
We note that the large-$\ell$ asymptotics for the quasinormal mode sum used in~\cite{Dolan:2011fh} should, in principle, be compounded
with a similar analysis of the branch cut of the Green's function - see, e.g.,~\cite{CO}; using in Schwarzshild the large-$\ell$ asymptotics similarly
to here in PH would instead entail investigating a two-dimensional PDE in terms of the time and radial coordinates.

Finally, we note that the four-fold singularity structure does not apply to spacetime points with angular separation equal to $\pi$. As shown in Section IV-F, the singularity structure is two-fold: $\delta\to-\delta\to\delta\to\cdots$. This pattern has recently been observed numerically in Schwarzschild spacetime~\cite{galley}, as has a corresponding pattern when $\gamma=0$, but with $PV(1/\sigma)\to-PV(1/\sigma)\to\cdots$. We have not considered this latter case, but it should be possible to do so using the Kirchhoff integral method. The presence of points with a two-fold singularity structure is also predicted in~\cite{Harte:2012uw}.


\begin{acknowledgments}
We thank Sam Dolan, Adrian Ottewill and Barry Wardell for interesting discussions.
M.C. acknowledges funding support from a IRCSET-Marie Curie International Mobility Fellowship in Science, Engineering and Technology. B.N. thanks Luis Lehner and Perimeter Institute for financial support.
\end{acknowledgments}


\appendix

\section{Distributional Limits}

\subsection{Derivation of (\ref{del1lim}).}
We have
\[ \Delta_1^{(\ep)}(z)=\frac{\sqrt{2}}{z}H(-z-2\ep)+\frac{\sqrt{2}}{z}\left(1-\frac{\sqrt{2\ep}}{\sqrt{z+2\ep}}\right)H(z+2\ep).\]
Let $\phi\in C^\infty_0(\mathbb{R})$. Then
\begin{eqnarray*}
<\Delta_1^{(\ep)},\phi> &=& \sqrt{2}\int_{\mathbb{R}\setminus[-2\ep,2\ep]}\frac{\phi(z)}{z}dz\\
&&+ \sqrt{2}\int_{-2\ep}^{2\ep}\left(\frac1z-\frac{\sqrt{2\ep}}{z\sqrt{z+2\ep}}\right)\phi(z)dz\\
&&- 2\sqrt{\ep}\int_{2\ep}^\infty\frac{\phi(z)}{z\sqrt{z+2\ep}}dz\\
&=:& <\Delta_{1a}^{(\ep)},\phi>+<\Delta_{1b}^{(\ep)},\phi>+<\Delta_{1c}^{(\ep)},\phi>.
\end{eqnarray*}
with the obvious term-by-term definitions of $\Delta_{1i}^{(\ep)}$, $i=a,b,c$.
By definition, we see that $\lim_{\ep\to0^+}\Delta_{1a}^{(\ep)}(z)=\sqrt{2}PV\left(\frac1z\right)$.
Substituting $z=:2\ep y$, we see that
\[ <\Delta_{1b}^{(\ep)},\phi> = \sqrt{2}\int_{-1}^1\frac{\sqrt{y+1}-1}{y\sqrt{y+1}}\phi(2\ep y) dy.\]
Then
\begin{eqnarray*} \lim_{\ep\to0^+}<\Delta_{1b}^{(\ep)},\phi> &=& \sqrt{2}\phi(0)\int_{-1}^1\frac{\sqrt{y+1}-1}{y\sqrt{y+1}} dy\\
&=&2\sqrt{2}\ln(1+\sqrt{2})\phi(0).\end{eqnarray*}
It follows that
\[ \lim_{\ep\to0^+}\Delta_{1b}^{(\ep)}=2\sqrt{2}\ln(1+\sqrt{2})\delta.\]

Likewise,
\[ <\Delta_{1c}^{(\ep)},\phi> =-\sqrt{2}\int_1^\infty \frac{\phi(2\ep y)}{y\sqrt{1+y}}dy,\]
so that
\begin{eqnarray*} \lim_{\ep\to0^+}<\Delta_{1c}^{(\ep)},\phi> &=& -\sqrt{2}\phi(0)\int_1^\infty\frac{dy}{y\sqrt{y+1}} dy\\
&=&-2\sqrt{2}\ln(1+\sqrt{2})\phi(0).\end{eqnarray*}
Thus
\[ \lim_{\ep\to0^+}\Delta_{1c}^{(\ep)}=-2\sqrt{2}\ln(1+\sqrt{2})\delta.\]

We conclude that
\[ \lim_{\ep\to0^+}\Delta_1^{(\ep)}(z) = \sqrt{2}PV\left(\frac{1}{z}\right),\]
which is (\ref{del1lim}).

\subsection{Derivation of (\ref{heplim}).}
Define
\[ \Delta_2^{(\ep)}(z) := \ln(\sqrt{z+2\ep}+\sqrt{2\ep})H(z+2\ep),\quad \Delta_2(z):=\ln(\sqrt{z})H(z).\]
We aim to prove the distributional limit
\[ \lim_{\ep\to0^+}\Delta_2^{(\ep)}=\Delta_2,\]
which is (\ref{heplim}). So let $\phi\in C^\infty_0(\mathbb{R})$. Then
\begin{eqnarray} <\Delta_2^{(\ep)}-\Delta_2,\phi> &=& \int_{-2\ep}^\infty\ln(\sqrt{z+2\ep}+\sqrt{2\ep})\phi(z)dz - \int_0^\infty\ln(\sqrt{z})\phi(z)dz \nn\\
&=& \int_{-2\ep}^0\ln(\sqrt{z+2\ep}+\sqrt{2\ep})\phi(z)dz+\int_0^\infty \ln\left(\frac{\sqrt{z+2\ep}+\sqrt{2\ep}}{\sqrt{z}}\right)\phi(z)dz.\label{heplimeq1}
\end{eqnarray}

With the substitution $z=:2\ep y$, the first integral here evaluates to
\[ 2\ep\int_{-1}^0\ln(1+\sqrt{1+y})\phi(2\ep y)dy + \ep\ln(2\ep)\int_{-1}^0\phi(2\ep y) dy,\]
which, for any test function $\phi$, vanishes in the limit $\ep\to 0^+$. Note that
\[ \lim_{\ep\to0^+}\int_{-1}^0\ln(1+\sqrt{1+y})\phi(2\ep y)dy=\frac12\phi(0).\]

Making the same substitution in the second integral on the right hand side of (\ref{heplimeq1}) yields
\[ \int_0^\infty \ln\left(\frac{\sqrt{z+2\ep}+\sqrt{2\ep}}{\sqrt{z}}\right)\phi(z)dz = 2\ep\int_0^\infty \ln\left(\frac{1+\sqrt{1+y}}{\sqrt{y}}\right)\phi(2\ep y)dy.\]
Since $\phi\in C^\infty_0(\mathbb{R})$, there are positive constants $C_\phi$ and $M_\phi$ such that
\[ \phi(z)=0 \hbox{ for all } z>2C_\phi\]
and
\[ |\phi(z)|\leq M_\phi \hbox{ for all } z\in\mathbb{R}.\]
Hence
\begin{eqnarray} \left|\int_0^\infty\ln\left(\frac{1+\sqrt{1+y}}{\sqrt{y}}\right)\phi(2\ep y)dy\right|&\leq& M_\phi\int_0^{{C_\phi}/2}\ln\left(\frac{1+\sqrt{1+y}}{\sqrt{y}}\right)dy\nn\\
&=&M_\phi\left\{\frac{\sqrt{C_\phi+\ep}}{\sqrt{\ep}}+\frac{C_\phi}{\ep}\ln\left(\frac{\sqrt{\ep}+\sqrt{C_\phi+\ep}}{\sqrt{C_\phi}}\right)\right\}.
\end{eqnarray}
It follows that
\[ \lim_{\ep\to0^+} \ep\int_0^\infty\ln\left(\frac{1+\sqrt{1+y}}{\sqrt{y}}\right)\phi(2\ep y)dy=0.\]
Hence for any test function $\phi$,
\[ \lim_{\ep\to0^+} <\Delta_2^{(\ep)}-\Delta_2,\phi>=0,\]
which proves (\ref{heplim}).

\section{Distributional forms of some infinite series}

\subsection{Derivation of (\ref{caform}).}

By standard Fourier transform methods, we have
\[ \sum_{\ell=1}^\infty \ell^{-1}\sin(\ell s))=\frac12(\pi-s),\qquad s\in(0,2\pi).\]
Thus the series converges on $\mathbb{R}$ to $\cu(s)$, the $2\pi$-periodic continuation to the real line of
\[ u(s) = \left\{ \begin{array}{cl}
                          0, & \hbox{if }s=0,\pi; \\
                          \frac12(\pi-s), & \hbox{if }0<s<2\pi.
                        \end{array}\right.\]
This function can be written as
\[ \cu(s) = \frac12(\pi-s)+\pi\left\{\sum_{k=1}^\infty H(s-2k\pi)-\sum_{k=0}^\infty H(-s-2k\pi)\right\}.\]
Similarly,
\[ \sum_{\ell=1}^\infty \ell^{-1}\cos(\ell s)=\frac12\ln\left(\frac{1}{2(1-\cos s)}\right),\qquad s\in(0,2\pi).\]
See for example Eq.1.441(1) of~\cite{GradRyz}. Periodicity of both sides indicates that the series converges almost everywhere on $\mathbb{R}$ to the locally integrable function of the right hand side defined on $\mathbb{R}\setminus2\pi\mathbb{Z}$. The series is therefore a distribution which we can identify with $\frac12\cd(s)$ where
\begin{eqnarray*}
\cd(s) &=& \ln\left(\frac{1}{2(1-\cos s)}\right)\\
&=& -2\ln|2\sin(s/2)|\\
&=& -2\left(\ln|s|+\sum_{k=1}^\infty|1-\frac{s^2}{4k^2\pi^2}|\right).
\end{eqnarray*}
This arises from elementary properties of logarithms and trigonometric functions and the infinite product representation
\[ \sin x = x\prod_{k=1}^\infty\left(1-\frac{x^2}{k^2\pi^2}\right).\]
Combining these results yields (\ref{caform}).

\subsection{Derivation of (\ref{cbform}).}

Formal differentiation of both sides of (\ref{caform}) yields (\ref{cbform}): the key point is that differentiating the right hand side of (\ref{caform}) involves the calculation of the derivative of a distribution which always yields another distribution. Thus (\ref{cbform}) is a valid distributional identity. The results (\ref{cdprimeform}) and (\ref{cuprimeform}) follow by applying the distributional derivatives
\[ \frac{d}{ds}\left\{\ln|s|\right\} = PV\left(\frac{1}{s}\right),\qquad \frac{d}{ds}\left\{H(s)\right\} = \delta(s).\] Some rearrangements of the series are also required.



\section*{References}



\end{document}